\documentclass[aps,prd,twocolumn,superscriptaddress,groupedaddress]{revtex4}
\usepackage{graphicx}  
\usepackage{dcolumn}   
\usepackage{bm}        
\usepackage{amssymb}   
\usepackage{amsmath}
\usepackage{color}
\usepackage{acronym}
\usepackage{algcompatible}
\usepackage{newfloat}
\usepackage{cancel}
\usepackage{hyperref}
\usepackage{caption}
\usepackage{subcaption}
\usepackage{changepage}
\graphicspath{{./figures/}}
\usepackage[T1]{fontenc}
\usepackage{svg}

\newcommand{\dd}{\mathrm d}

\begin{document}
\title{Rapid parameter estimation for an all-sky continuous gravitational wave search using conditional varitational auto-encoders}

\author{Joe Bayley}
\author{Chris Messenger}
\author{Graham Woan}
\affiliation{SUPA, University of Glasgow, Glasgow G12 8QQ, United Kingdom.}


\begin{abstract}
All-sky searches for continuous gravitational waves are generally model dependent and computationally costly to run. By contrast, SOAP is a model-agnostic search that rapidly returns candidate signal tracks in the time-frequency plane. In this work we extend the SOAP search to return broad Bayesian posteriors on the astrophysical parameters of a specific signal model. These constraints  drastically reduce the volume of parameter space that any follow-up search needs to explore, so increasing the speed at which candidates can be identified and confirmed. Our method uses a machine learning technique, specifically a conditional variational auto-encoder, and delivers a rapid estimation of the posterior distribution of the four Doppler parameters of a continuous wave signal. It does so without requiring a clear definition of a likelihood function, or being shown any true Bayesian posteriors in training. We demonstrate how the Doppler parameter space volume can be reduced by a factor of $\mathcal{O}(10^{-7})$ for signals of SNR 100.
\end{abstract}

\maketitle

\acrodef{GW}[GW]{gravitational wave}
\acrodef{CW}[CW]{continuous gravitational wave}
\acrodef{NS}[NS]{neutron star}
\acrodef{EM}[EM]{electromagnetic}
\acrodef{SNR}[SNR]{signal-to-noise-ratio}
\acrodef{LIGO}[LIGO]{Laser Interferometer Gravitational-wave Observatory}
\acrodef{SFT}[SFT]{short Fourier transform}
\acrodef{FFT}[FFT]{fast Fourier transform}
\acrodef{UCD}[UCD]{up, centre or down}
\acrodef{MDC}[MDC]{mock data challenge}
\acrodef{PSD}[PSD]{power spectral density}
\acrodef{ROC}[ROC]{receiver operating characteristic}
\acrodef{RMS}[RMS]{root median square}
\acrodef{MCMC}[MCMC]{Markov-Chain Monte Carlo}
\acrodef{CNN}[CNN]{convolutional neural network}
\acrodef{CBC}[CBC]{compact binary coalescence}
\acrodef{RELU}[RELU]{rectified linear unit}
\acrodef{vitstat}[vitstat]{Viterbi statistic}
\acrodef{vitmap}[vitmap]{Viterbi map}
\acrodef{CPU}[CPU]{central processing unit}
\acrodef{GPU}[GPU]{graphics processing unit}
\acrodef{CVAE}[CVAE]{conditional variational auto-encoder}
\acrodef{Neville}{NEutron star Variational Inference with a Lack of Likelihood Evaluation}

\section{\label{intro:1}Introduction}
%

%
Non-axisymmetric and rapidly rotating neutron stars are expected to produce detectable \acp{GW} in the sensitive frequency range of ground based detectors such as \ac{LIGO} \cite{LIGOScientific:2014pky} and Virgo \cite{VIRGO:2014yos}. 
They would be seen as long-duration quasi-sinusoidal signals. 
A number of specific mechanisms have been proposed for this emission, including r-mode oscillations and elastic or magnetic deformations to the crust of the neutron star (see \cite{sieniawska2019ContinuousGravitationala, owen2009ProbingNeutrona} for a review). 
Such observations would provide new insights into neutron star physics, including constraints on the equation of state of hot, dense matter.

Searches for these types of \acp{CW} generally fall into three categories, based on the assumptions made about the source and signal prior to the search. 
Targeted searches \cite{dupuis2005BayesianEstimation, schutz1998DataAnalysis, TargetedO3} use electromagnetic observations to provide information on the sky location, frequency, and frequency derivatives of signals from known pulsars. 
Directed searches \cite{piccinni2020DirectedSearch, DirectedSNRO3, DirectedSCOX1O3, DirectedMilkyO3, DirectedCasAO3} use electromagnetic observations to provide information on the sky location only, and all-sky searches \cite{AllSkyBinaryO3, AllSkyIsolatedearlyO3, AllSkyIsolatedO3, RodrigoBinaryAllsky} explore all sky locations and a broad span  of frequency, and frequency derivative parameter space.  In this paper we will concentrate on this final category of search.

%
All-sky searches probe a very large parameter volume, and it is not computationally feasible to apply the fully-coherent matched filtering technique used by targeted searches~\cite{dupuis2005BayesianEstimation, schutz1998DataAnalysis, TargetedO3} in this regime. Instead one can use a semi-coherent approach, in which the data is divided into segments which are analysed separately. The coherent analysis of each segment is then  incoherently combined using various techniques \cite{FreqHough, SkyHough, JKS,AllskyO1}, see \cite{Tenorio:cwreview} for a review. In general, as the length of the segments (i.e., the coherence length) increases, the sensitivity of the search also increases but at a computational cost. Semi-coherent methods are designed to balance this computational cost against the sensitivity of the search.

%
One of the fastest all-sky search methods for \acp{CW} is SOAP~\cite{bayley2019SOAPGeneralised}. SOAP performs a search for weakly-modelled signals, without a specific astrophysical justification, and therefore explores the entire parameter space that might contain a \ac{CW} signal as well as signals that do not follow the standard \ac{CW} frequency evolution. This search is explained in more detail in Sec.~\ref{soap} and in \cite{bayley2019SOAPGeneralised, bayleySOAPCNN, bayley2020Soapcw}.

%
SOAP was designed to identify signal candidates rapidly so that they could be followed-up later by more sensitive parameter-dependent methods. Once SOAP identifies a signal it therefore needs to give estimates of the candidate's frequency parameters and sky position to these follow-up searches. 
The outputs from the SOAP search include the frequency bin location as a function of time for a candidate, producing tracks which can potentially randomly wander through the frequency band.
The difficulty in defining a clear likelihood for these tracks means that we cannot use traditional methods to produce Bayesian posterior distributions. 
In this work we turn to likelihood free methods \cite{sbi_frontier, gabbard2019BayesianParameter} and introduce our implementation named Neville which leverages machine learning to generate Bayesian posteriors on the four Doppler parameters of the \ac{CW} signal: the sky position $\alpha, \delta$, the frequency $f_0$, and the frequency derivative $\dot{f}_0$. 

In Sec.~\ref{soap} we introduce the SOAP method and some of the key outputs from the search as well as the standard model that is used for a \ac{CW} signal. 
In Sec.~\ref{machine} we introduce how machine learning has been used for Bayesian parameter estimation, in particular we describe our \ac{CVAE} implementation known as Neville and how it can be trained to approximate a Bayesian posterior.
In Sec.~\ref{data} we discuss the different data-sets which are generated for the training and testing of the method described in Sec~\ref{machine}, and introduce different parameterisations of the astrophysical parameters. 
In Sec.~\ref{networkdesign} we outline the specifics of the \ac{CVAE} model and its structure and then describe the training procedure in Sec.~\ref{training} and the timing in Sec.~\ref{timing}.
In Sec.~\ref{results} we show the results from testing this method on the two data-sets described in Sec.~\ref{data} and discuss how this method is used in practice.

\section{\label{soap} SOAP }
%

%
SOAP~\cite{bayley2019SOAPGeneralised, bayleySOAPCNN, bayley2020Soapcw} is a search pipeline for weakly-modelled
long-duration signals, based on the Viterbi
algorithm~\cite{viterbi1967ErrorBounds}. In its simplest form SOAP
analyses a spectrogram to find the continuous time-frequency track that contains the greatest total spectral power. If a signal is present and sufficiently strong, then this track is likely to follow its frequency evolution very closely.
In~\cite{bayley2019SOAPGeneralised} the SOAP algorithm was expanded to include
multiple detectors as well as a statistic to penalise instrumental artefacts in the data. This was followed by further developments in~\cite{bayleySOAPCNN} where convolutional neural networks were used on the outputs of SOAP to improve the robustness of the search against instrumental artefacts.

\begin{figure*}
	\includegraphics[scale=0.48]{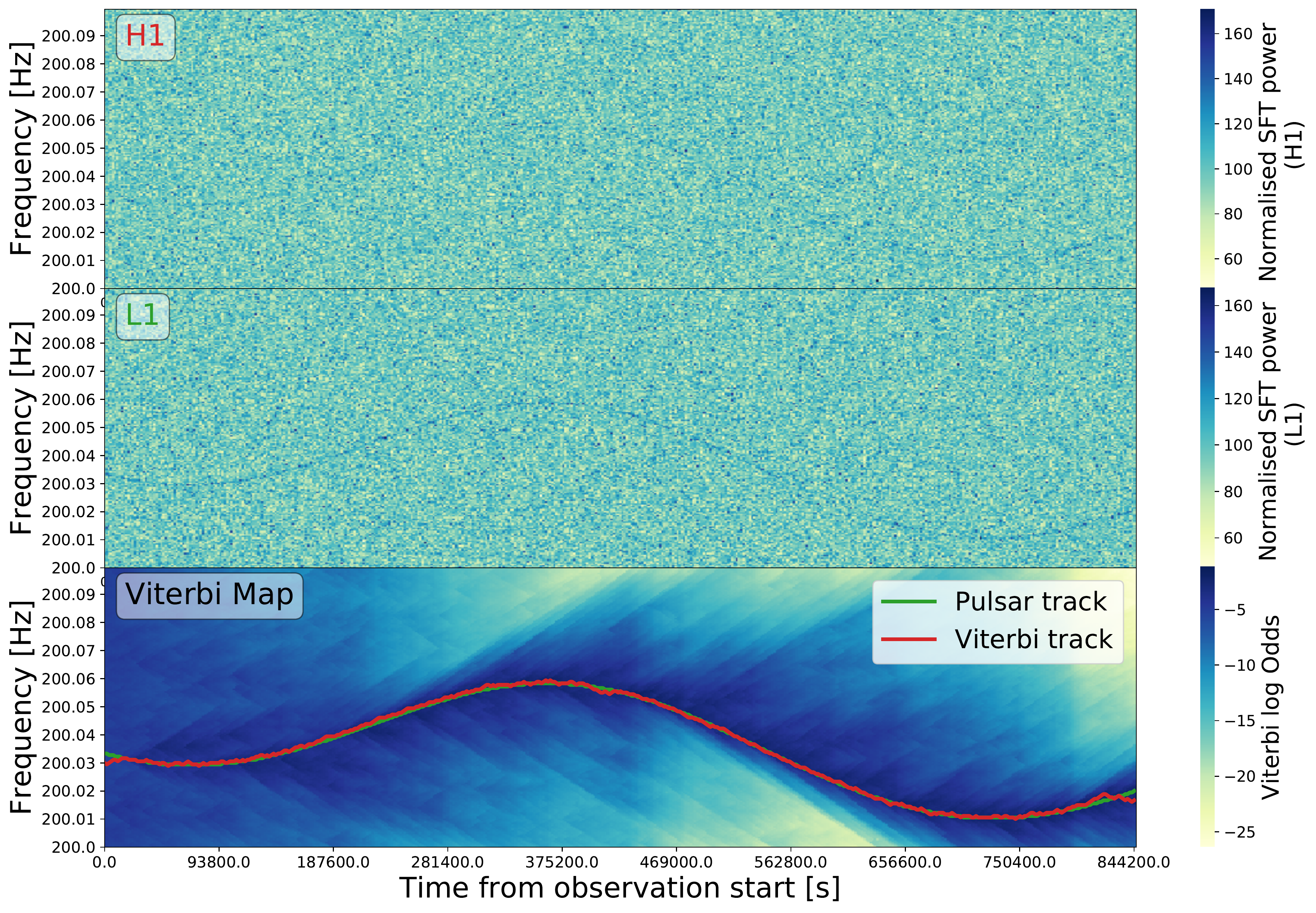}
	\caption{\label{soap:viterbiplot} The top two
		panels show simulated time-frequency spectrograms from the \ac{LIGO} Hanford and Livingston observatories \cite{LIGOScientific:2014pky}. The data includes a simulated \ac{CW} signal with an optimal network \ac{SNR} of 120. The bottom panel shows the normalised Viterbi map with the pixel intensity showing the Viterbi log-odds that the track falls in a particular frequency bin as a function of time. The green curve shows the injected frequency evolution of the signal and the red line shows the recovered track from SOAP.}
\end{figure*}

%
An example of the inputs and outputs of the SOAP algorithm is shown in Fig.~\ref{soap:viterbiplot}. The figure shows the input time-frequency spectrograms and the three main output components: the Viterbi track, the Viterbi statistic and the Viterbi map, described below.

%
\begin{description}
	\item [Viterbi track] The Viterbi track is the most probable track through time-frequency data given a choice of statistic (i.e., summed \ac{SFT} power).
	\item [Viterbi statistic] The Viterbi statistic is the sum of the
	individual statistics along the Viterbi track. In the analysis that
	follows we use the `line-aware' Viterbi statistic. This is the sum of
	the log-odds ratios, $p_{\mathrm signal}/(p_{\mathrm line} + p_{\mathrm noise})$
	along the track \cite{bayley2019SOAPGeneralised}.
	\item [Viterbi map] The Viterbi map shows the value of the Viterbi
	statistic for every time-frequency bin in the spectrogram,
	corresponding to the log-probability that the track passes through each
	time-frequency bin. Each time slice in the map is normalised
	individually, i.e., each vertical slice is adjusted so that the sum of
	their exponentiated values is unity. Each pixel in the image can
	therefore be interpreted as a value related to the log-probability that
	the signal has a particular frequency conditioned on the time of the vertical slice.

\end{description}

Using the techniques described in~\cite{bayley2019SOAPGeneralised, bayley2020Soapcw},  the only relevant information for later investigations provided from the track is the narrow frequency band (0.1 Hz) in which the signal was found. Although useful, this still leaves a large parameter volume for a follow-up search to explore. In this paper we describe how the Viterbi track, and therefore the potential frequency evolution of a source, can be used to infer the \ac{CW} Doppler parameters: the frequency $f_0$, its derivative $\dot{f}_0$, and the sky location $(\alpha, \delta)$.  This process is complicated by the unusual statistical properties of the Viterbi track, and its non-stationary deviations from a the true Doppler-modulated signal shape in the presence of noise.
\subsection{\label{cw}Continuous Wave Signal}

In the frame of the source, \ac{CW} signals are usually modelled as a quasi-sinusoidal, with a slow frequency evolution over time due (for example) to radiative losses. A ground based detector will see these signals modulated in amplitude, due to the detector antenna pattern, and Doppler-modulated in frequency due to the non-inertial motion of both the source and the detector. In the standard SOAP search, strips of constant time in the time-frequency plane contain the mean of 30-minute spectra over a day, so the modulation from the Earth's spin and  from the antenna pattern are not apparent. Relativistic effects are small at these resolutions, so the frequency evolution of the signal $f(t)$ is simply
\begin{equation}
	\label{cw:freqevolution}
	f(t) = \frac{1}{2\pi} \frac{\dd \Phi(t)}{\dd t} = f_0(t)\left( 1 + \frac{\bm{v}(t) \cdot \hat{\bm{n}}}{c}\right),
\end{equation}
where $\Phi(t)$ is the phase evolution of the signal, $\bm{v}(t)$ is the Earth's velocity relative to the source, $\hat{\bm{n}}$ is the unit vector pointing towards the source, and $c$ is the speed of light. The signal frequency $f_0(t)$ seen in the solar system's barycentric frame is usually represented by a Talor expansion,
\begin{equation}
	f_0(t) = f_0 + \dot{f_0}(t-t_0) + \ldots ,
\end{equation}
and in this work we will concentrate on the first two terms in this expansion. 
The velocity $\bm{v}(t)$ of the earth relative to any object at any given time $t$ is defined using solar system ephemerides data via the \texttt{lalsuite} library \cite{lalsuite}.

The full frequency evolution of the observed signal then depends on the source's barycentric frequency $f_0$ and its derivative $\dot{f}_0$, and the sky position $\alpha, \delta$.  Given such a signal in Gaussian noise one could determine the joint posterior probability distribution of these parameters using standard Bayesian sampling techniques such as MCMC or nested sampling. However, extracting parameters from a Viterbi track, as returned by SOAP, is less straightforward. 

\begin{figure}
	\includegraphics[width=\linewidth]{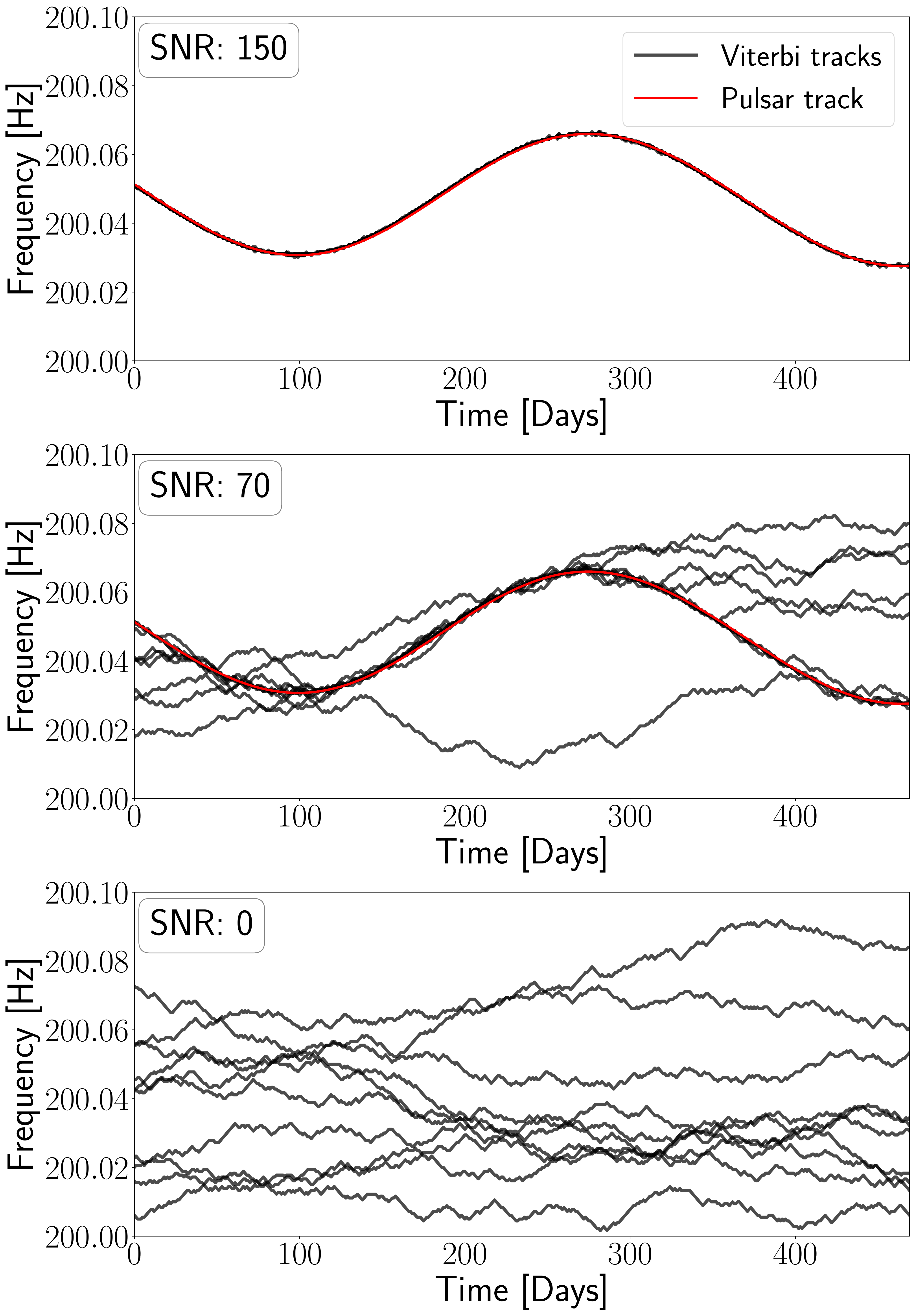}
	\caption{\label{soap:vittracks} Each of the panels show an example of many Viterbi tracks (black) each resulting from running SOAP on the same simulated signals with different noise realisations over a 460 day duration. The red curve shows the true \ac{CW} frequency evolution. The top panel shows a high \ac{SNR} of 150, the middle panel shows a signal with \ac{SNR} 70 which is at the edge of our detection threshold and the lower panel shows Viterbi tracks resulting from just noise realisations.}
\end{figure}

If the \ac{SNR} is large enough, the Viterbi track will closely follow the frequency evolution of a signal in the time-frequency plane. 
If the \ac{SNR} is very low the Viterbi algorithm will simply follow noise and the track will wander stochastically.  
Between these extremes of SNR we see both behaviours: tracks that spend some of the time locked to the signal and some time tracking noise, examples of which can be seen in Fig.~\ref{soap:vittracks}. 
The equivalent noise in these tracks is highly correlated, non-stationary and non-Gaussian, making it is difficult to write down a corresponding likelihood function.

Due to there being no clear way to calculate the likelihood, traditional sampling methods cannot be used for this particular problem.
In this work we therefore look to using likelihood free methods to extract the Bayesian posteriors \cite{sbi_frontier}, in particular we used a form of \ac{CVAE} which is explained in more detail in Sec.~\ref{machine}. 
This machine learning based method, allows us to extract Bayesian posteriors without ever being trained on the true posteriors or defining a likelihood function.

\section{\label{machine} Machine learning and parameter estimation}
%
Within the field of \acp{GW} the use of machine learning is becoming more prevalent \cite{EnhancingGWs} with methods being developed for many tasks including detection and inference.
In particular, a number of methods have been developed to estimate the Bayesian posterior distribution on the parameters of \acp{CBC}  using machine learning, including the use of \acp{CVAE} \cite{gabbard2019BayesianParameter} and normalising flows  \cite{green2020GravitationalwaveParameter}. 
In the following work we apply a \ac{CVAE} to estimate the posterior probability distributions of the parameters considered in the preceding section. 
Using this \ac{CVAE} implementation one can estimate the Bayesian posterior without explicitly being shown the true posterior or likelihood during the training procedure. 
Only the prior parameter space and noise model are assumed, and the data used to train the \ac{CVAE} is drawn from this parameter space.
In the case of Viterbi tracks, we can write down a prior parameter space for the signal parameters, but as we cannot write down a noise model we either numerically simulate noise instances or take examples from real data.

The objective of the \ac{CVAE} is to minimise the cross-entropy between the Bayesian posterior $p(x | y)$ and a target distribution $r_{\theta}(x | y)$ described by neural network parameters $\theta$. The following section follows the derivations in \cite{gabbard2019BayesianParameter}. The cross-entropy is defined as
\begin{equation}
	\label{cvae:crossentropy}
	H(p,r) = -\int p(x | y) \log r_{\theta}(x | y) \,\dd x,
\end{equation}
where $x$ are the parameters of the model, $y$ is the data and $\theta$ are the learned parameters of the neural network describing the distribution $r_{\theta}(x|y)$. This cross-entropy is minimised when $p(x | y) = r_{\theta}(x | y)$.
However, the cross-entropy cannot be calculated directly for every training example since computing the Bayesian posterior $p(x | y)$ is costly or as in our case we have no clear definition.
We can instead minimise the expectation value of the cross-entropy over the distribution of instances of $y$, which would make the distributions as similar as possible over all possible $y$. 
The expectation value of the integral can be written as
\begin{equation}
	\begin{split}
	\langle H \rangle &= -\iint p(y)p(x | y) \log r_{\theta}(x | y)  \,\dd x \,\dd y\\
	&= -\iint p(x)p(y | x) \log r_{\theta}(x | y) \,\dd x \,\dd y,
	\end{split}
\end{equation}
where $p(x)$ is the prior distribution on the parameters. Hence we are now taking the expectation over both the noise realisation and the signal parameters. The target distribution $r_{\theta}(x | y)$ can be parametrised as a combination of two distributions known as an encoder $r_{\theta_1}(z | y)$ and a decoder $r_{\theta_2}(x | y, z)$ described by a neural network with parameters $\theta_1, \theta_2$. 
By marginalising over this latent space we can write the target distribution as
\begin{equation}
	\label{cvae:targetdist}
	r_{\theta}(x | y) = \int r_{\theta_1}(z | y) r_{\theta_2}(x | y, z) \,\dd z,
\end{equation}
where the $r_{\theta_1}$ defines a probability distribution in the latent space $z$ and $r_{\theta_2}$ describes a distribution in the physical parameter space $x$ and is conditional on the data and latent space location.
The latent space $z$ is an abstract representation of the input which is learned by the \ac{CVAE} and $\theta_1$ and $\theta_2$ represent the trainable parameters of the neural networks.
After some manipulation, shown in~\cite{gabbard2019BayesianParameter}, and the addition of a second encoder network $q_{\phi}(z | x, y)$ which depends on both the measurement $y$ and parameters $x$, one finds that the expectation of the cross entropy satisfies
\begin{equation}
    \label{cvae:costint}
	\begin{split}
	\langle H \rangle \leq - \int \int p(x) p(y | x) \mathrm{E}_{q_{\phi}(z | x, y)}\Big\{ \log{r_{\theta_2}(x | y, z)}  \\ 
	 - \mathrm{KL}\left[q_{\phi}(z | x, y) || r_{\theta}(x | y) \right] \Big\} \,\dd y \, \dd x,
	\end{split}
\end{equation}
where $\mathrm{KL}$ is the KL divergence and $\mathrm{E}_{q_{\phi}(z | x, y)}$ is the expectation value over the distribution of $q$.
This integral can be approximated via Monte-Carlo integration where samples of $x$ and $y$ are drawn from the prior $p(x)$ and the likelihood $p(y | x)$. This allows it to be used as the cost function to be minimised in the training of the 3 neural networks modelling the $r_{\theta_1},r_{\theta_2}$ and $q_{\phi}$ distributions. 
The three neural networks model these distributions by outputting parameters describing a distribution, i.e. the mean and variance of a Gaussian distribution. 
The cost function then approximates Eq.~\ref{cvae:costint} by taking the average over a batch or draws from the prior and likelihood such that 
\begin{equation}
	\begin{split}
		\label{cvae:cost}
		\langle H \rangle \lesssim \frac{1}{N_b} \sum_{n=1} ^{N_b}\left[ -\log r_{\theta_2}(x_n | z_n, y_n) \right. \\ \left. + \mathrm{KL}\left[q_{\phi}(z_n| x_n, y_n) || r_{\theta_1}(z_n| y_n) \right] \right]
	\end{split}
\end{equation}
where $N_b$ is the number of instances of $x$ and $y$ used per training step (the batch size). 
The right hand side of Eq.~\ref{cvae:cost} is then used to train the network.

\subsection{Training \label{ml:training}}

The aim of the training procedure is to adjust the network parameters to minimise the cost function described by Eq.~\ref{cvae:cost} and therefore also Eq.~\ref{cvae:crossentropy}. To do this we calculate two main components: the `reconstruction loss' $\log r_{\theta_2}(x_n | z_n, y_n)$ and the KL divergence $\mathrm{KL}\left[q_{\phi}(z_n| x_n, y_n) || r_{\theta_1}(z_n| y_n) \right]$.
\begin{enumerate}
\item To calculate the reconstruction loss, first the data $y$ and parameters $x$ are propagated through the $q_{\phi}(z| x, y)$ encoder which outputs parameters describing the latent space $z$. 
These are a set of means $\mu_{\phi}$ and variances standard deviations $\sigma_{\phi}$ of an uncorrelated Gaussian distribution with $n_{z}$ dimensions. 
One can then sample from this Gaussian distribution and combine the outputs $z$ with the data $y$ and propagate this through the $r_{\theta_2}(x | z, y)$ decoder which outputs the means $\mu_{x}$ and standard deviations $\sigma_{x}$ of another uncorrelated Gaussian distribution with $n_x$ dimensions. 
The reconstruction loss can then be calculated by evaluating the output Gaussian distribution at the true values of $x$.
\item The KL divergence is calculated using the outputs of the $q_{\phi}(z| x, y)$ encoder and the $r_{\theta_1}(z| y)$ encoder. The $r_{\theta_1}(z| y)$ encoder takes the data $y$ as input and also outputs the means $\mu_{\theta_1}$ and standard deviations $\sigma_{\theta_1}$ of an uncorrelated Gaussian distribution with $n_z$ dimensions. 
There is no analytical form for the KL divergence between multivariate Gaussians, but as in \cite{gabbard2019BayesianParameter} we can approximate it as 
\begin{equation}
    \mathrm{KL}\left[ q_{\phi}(z| x, y) || r_{\theta_1}(z| y)\right] \approx \left. \log \left(\frac{q_{\phi}(z| x, y)}{r_{\theta_1}(z| y)}  \right) \right|_{z \sim q_{\phi}(z| x, y)}.
\end{equation}
This is a single sample estimate of the KL divergence, therefore the average of these values is then taken over a batch.

\end{enumerate}

The reconstruction loss and the KL divergence are combined to form the cost function as in Eq.~\ref{cvae:cost}, where the expectation is estimated over a batch of input training data of size $N_b$.
This cost function is then minimised over many batches using back-propagation, where the ADAM optimizer~\cite{adamopt} is used with the default parameters. 
During training, we modify a weight on the KL and reconstruction loss components where we do not optimise the entire loss function at once. 
Initially we optimise the reconstruction loss and slowly introduce the KL divergence term into the calculation with a multiplicative pre-factor. 
The pre-factor linearly increases from 0 to 1 over 300 epochs avoiding the known local minima where the KL term remains close to zero and the latent space structure is not learnt by the $r_{\theta_1}$ and $q_{\phi}$ networks.

\subsection{Testing \label{machine:testing}}

When generating samples from the posterior estimate, the procedure is slightly different to training. 
The aim here is to perform the integral in Eq.~\ref{cvae:targetdist} using Monte Carlo integration which can be written as
\begin{equation}
    r_{\theta}(x|y) \propto \sum_{i}^{N}{r_{\theta_2}(x|y, z_i)|_{z_i \sim r_{\theta_1}(z | y)}},
\end{equation}
where $N$ is the number of samples. We do not perform this directly however, but generate samples of $x$ from $r_{\theta}(x|y)$ by sampling from $r_{\theta_2}(x|y, z)$ conditional on $z$ samples drawn from $r_{\theta_1}(z|y)$:
\begin{equation}
    x \sim r_{\theta_2}(x|y, z)|_{z \sim r_{\theta_1}(z | y)}.
\end{equation}
To generate posterior samples we need to generate samples in the latent space $z$, now using the $r_{\theta_1}(z| y)$ encoder which takes input of only $y$. We can then make many draws from $r_{\theta_1}(z| y)$ described by $\mu_{\theta_1}$ and $\sigma_{\theta_1}$. 
These latent space $z$ samples can then each separately be fed into the decoder $r_{\theta_2}(x | z, y)$ along with the data $y$ to generate a set of means $\mu_{x}$ and standard deviations $\sigma_{x}$ of a Gaussian distribution. 
From each of these we can draw a single sample in the physical parameter space of $x$. 
It is these samples that we treat as being drawn from the posterior distribution.
It is important to note here, that whilst the output of $r_{\theta_1}(z| y)$ and $r_{\theta_2}(x | z, y)$ are an uncorrelated Gaussian distributions, this does not mean that the final distribution $r_{\theta}(x | y)$ is also Gaussian. 
The variation provided by the latent space distribution which is marginalised over in Eq.~\ref{cvae:targetdist} allows for a diverse family of possible output distributions in the physical space.

\section{Data \label{data}}

To follow the training procedure outlined in Sec.~\ref{ml:training} one needs many examples of the data $y$ and the corresponding parameters $x$.
Two distinct data-sets are used to test the \ac{CVAE} described in Sec.~\ref{machine}; both use measurements of frequency as a function of time as the input however each have different noise models. One data-set uses the \ac{CW} signals frequency bin location with a simplified noise model (Sec.~\ref{data:gaussnoise}) to allow comparison to standard techniques. The other data-set has the noise in the form of Viterbi tracks output from SOAP (Sec.~\ref{data:viterbinoise}) and is the main use case for this method.

\begin{table*}
	\caption{\label{data:sigpars} The upper and lower bounds for the random signal parameters. The parameters $\alpha$, $\sin\delta$, $f_0$, $\log\dot{f}_0$, $\cos\iota$, $\phi_0$, $\psi$ and \ac{SNR} were sampled uniformly between these bounds in each band. The frequencies
		$f_\mathrm{min}$ and $f_\mathrm{max}$ refer to the band limits, and signals are randomly placed in the centre half of the band. $f_\mathrm{min}$ is arranged on a uniform grid with a 0.1\,Hz spacing in the given range and the bandwidth $f_\mathrm{max} - f_\mathrm{min} = 0.1$\,Hz. Except for the distribution of signal
		frequencies $f_0$, all the injections parameters are sampled from the same
		distributions as the S6 \ac{MDC}~\cite{walsh2016ComparisonMethods}.}
	\bgroup
	\def\arraystretch{1.5}
	\centering
	\begin{tabular}{c c c c c c c c c c r|}
		\hline
		\hline
		& $\alpha$ [rad]& $\sin\delta$ [rad] & $f_0$ [Hz]&
		$\log_{10}\left(\dot{f}_0 [\mathrm{Hz/s}]\right)$ & $\cos{\iota}$ [rad]& $\phi$ [rad]& $\psi$ [rad] & SNR & $f_\mathrm{min}$ [Hz]\\
		\hline
		lower bound & $0$ & $-1$ & $f_\mathrm{min} + 0.25(f_\mathrm{max} - f_\mathrm{min})$ & $-9$ & $-1$ & 0 & 0 & 60 & 40\\
		\hline
		upper bound & $2\pi$ & $1$ & $f_\mathrm{min} + 0.75(f_\mathrm{max} - f_\mathrm{min})$ & $0$ & 1 & $2\pi$ & $\pi/2$ & 150 & 500\\
		\hline
	\end{tabular}
	\egroup
\end{table*}

\begin{figure}
	\includegraphics[width=\linewidth]{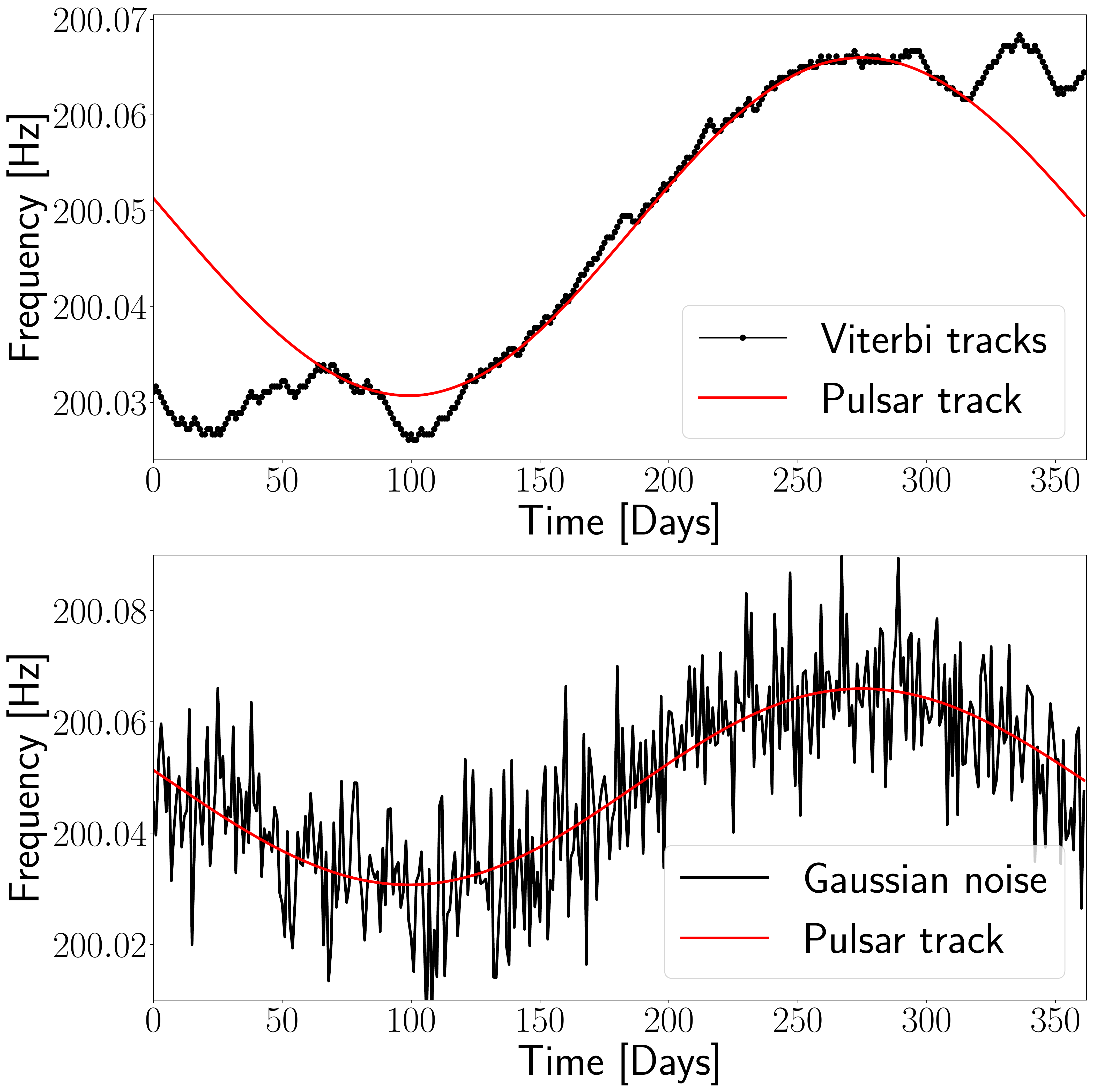}
	\caption{\label{data:vittracks} Examples of the two types of input data $y$ that are used to train the \ac{CVAE}. The top panel shows an example of a Viterbi track output from SOAP, overlaid is the true signal path in red. The lower panel shows the data-set where Gaussian noise has been added to the frequency components of the \ac{CW} signal as a function of time.}
\end{figure}

\subsubsection{Gaussian noise dataset \label{data:gaussnoise}}

The first data-set is generated by simulating a \ac{CW} signal using parameters drawn from the prior distribution described in Tab.~\ref{data:sigpars}.
The true instantaneous frequency of the signal can then be found at a given set of times which cover a time-span of 362 days sampled once per day.
This is chosen to have the same input size as the realistic case described in Sec.~\ref{data:viterbinoise}.
Once we have the frequency of the \ac{CW} signal over a range of times we add independent Gaussian noise samples with a mean of 0 and standard deviation of 0.01Hz to each of the frequency locations (arbitrarily chosen to be 1/10th of the band width). Whilst this is not a realistic noise distribution it allows for direct comparison to existing Bayesian sampling techniques and a way to validate the technique. An example of this type of input can be seen in the lower panel of Fig.~\ref{data:vittracks}. The data is then scaled to be between 0 and 1 using 
\begin{equation}
    \label{data:scaling}
    f_{\mathrm{scaled}}(t) = \frac{f(t) - f_{\mathrm{min}}}{f_{\mathrm{max}} - f_{\mathrm{min}}},
 \end{equation}
 where $f(t)$ is the frequency location as a function of time and $f_{\mathrm{min}}$ and $f_{\mathrm{max}}$ are the upper and lower edges of the analysis band defined in Tab.~\ref{data:sigpars}.
 In total we generate $10^6$ training signals in the 40-500 Hz range recording both their Doppler parameters and the scaled frequency track.

\subsubsection{Viterbi noise dataset \label{data:viterbinoise}}

The second data-set consists of frequency tracks with Viterbi noise where an example can be seen in the upper panel of Fig.~\ref{data:vittracks}. The Viterbi tracks are generated by running the SOAP search on a set of \ac{CW} simulations which have their parameters distributed according to the prior distribution described in Tab.~\ref{data:sigpars}, i.e., they are distributed and transformed in the same way as the previous test.
The \ac{SNR} defined in Tab.~\ref{data:sigpars} and Eq.15 of \cite{PrixSNR} is achieved by re-scaling the \ac{GW} amplitude $h_0$ based on the noise \ac{PSD}.
The power spectrum of the signal can then be simulated for each time segment of the spectrogram, this is done by assuming the time-series distributed according to Gaussian noise therefore producing a spectrogram which is $\chi^2$ distributed. 
The signal power will then be distributed according to a non-central $\chi^2$ distribution with a non-centrality parameter equal to the square of the \ac{SNR}.

The SOAP search is setup up similarly to in \cite{bayley2019SOAPGeneralised, bayleySOAPCNN}, where we use the line aware statistic \cite{bayley2019SOAPGeneralised} with parameters $w_{\mathrm{S}} = 4.0$, $w_{\mathrm{L}} = 10$ and $p(M_{\mathrm{L}})/p(M_{\mathrm{S}}) = 0.4$, where $w_{\mathrm{S}}$ is the prior width in \ac{SNR} of the signal model, $w_{\mathrm{L}}$ is the prior width in \ac{SNR} of the line model and $p(M_{\mathrm{L}})/p(M_{\mathrm{N}})$ is the prior odds ratio ratio for the signal and noise models.  
The Viterbi tracks output from SOAP are then scaled such that the analysis band is in the range of 0 to 1, as described in Eq.~\ref{data:scaling}.
The scaled Viterbi tracks are then what is used for the data $y$ in the \ac{CVAE}.

\begin{figure}
	\includegraphics[width=\linewidth]{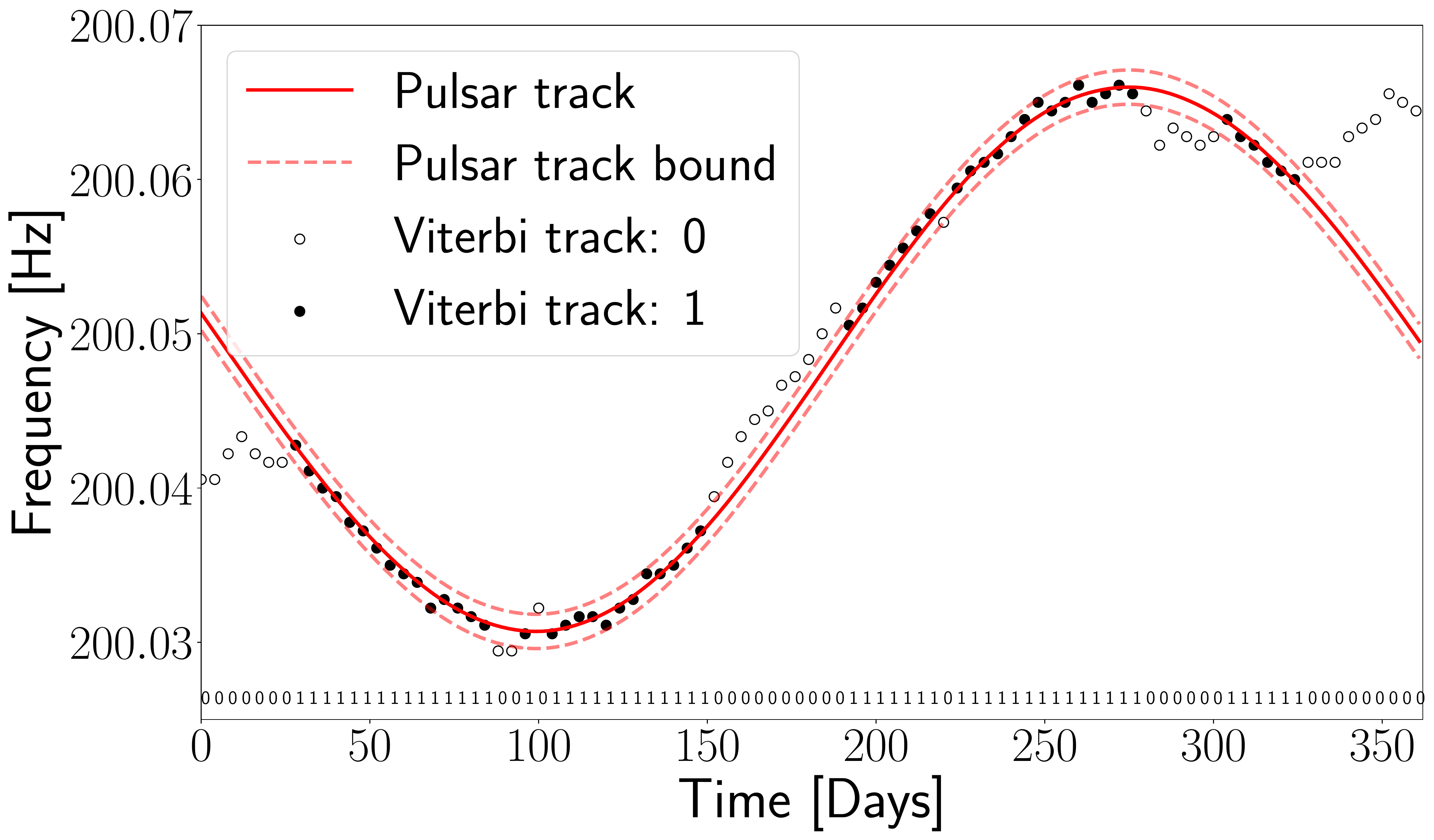}
	\caption{\label{data:booltracks} Examples of a Viterbi track and how it relates to each of the boolean values, the Viterbi track has been down-sampled to allow for easier viewing of the track elements. If the Viterbi track falls outside the pulsar tracks bound (red dashed) then it is assigned a value of 0 (unfilled circles) otherwise it is assigned a value of one (filled circles). The list of boolean values corresponding to the Viterbi track is shown at the bottom of the image.}
\end{figure}

The parameters $x$ for this particular \ac{CVAE} consist of the four Doppler parameters and an extra condition for each track element indicating whether is is associated with a signal or not. 
These conditions were introduced to provide extra information to help the \ac{CVAE} learn the Doppler posterior distributions more effectively.
The form of the conditions is in 362 boolean values which identify which of the track elements are within two frequency bin widths of the true signal. 
This value is chosen since power in frequency bins outside of this range is unlikely to be associated with the injected signal.
The boolean values $b$ are defined by
\begin{equation}
    \label{data:booleq}
    b(t) = 
    \begin{cases}
    1 & \text{for } |f_{\mathrm{viterbi}}(t) - f_{\mathrm{pulsar}}(t)| < \frac{2}{t_{\mathrm{SFT}}} \\
    0 & \text{otherwise}
    \end{cases},
\end{equation}
where $f_{\mathrm{viterbi}}$ is the frequency bin location of the viterbi track, $f_{\mathrm{pulsar}}$ is the frequency bin location of the pulsar signal and $t_{\mathrm{SFT}}$ is the length of a \ac{SFT}.
Figure \ref{data:booltracks} shows an example of a Viterbi track, the boundaries chosen around the true signal, and the results of applying Eq.~\ref{data:booleq}to these.

In total we simulate $10^6$ training examples in the 40-500 Hz range and generate the Doppler parameters, Viterbi tracks and boolean arrays for each.

\subsection{Parameterisation}
\label{data:parameterisation}

It is often useful to choose a different parameterisation of the signal in order to simplify the problem for the \ac{CVAE}, this can allow for faster training and better performance. 
In this example the signal has four Doppler parameters which we are interested in: the equatorial sky positions $\alpha$ and $\delta$ and the frequency $f_0$ and its derivative $\dot{f}_0$.There are two main transformations that are made before normalising the parameters between 0 and 1.

The first is to convert the equatorial sky positions $\alpha, \delta$ into the ecliptic longitude $\gamma$ and latitude $\lambda$.
The Viterbi tracks cannot be used to distinguish between the upper and lower ecliptic hemispheres as, due to only being sampled once per day, they have no access to daily Doppler or antenna pattern modulation.
By parameterising the sky position in the ecliptic frame our prior range only has to cover one hemisphere as this is duplicated in opposite hemisphere. The posterior should then contain only a single mode simplifying the problem for the \ac{CVAE}.

The second transformation is to convert the signal frequency $f_0$ into an offset from the lower edge of each analysis band $f_{\mathrm min}$.
This allows us to normalise the offset parameter between 0 and 1 rather than the entire 40-500 Hz frequency range, allowing the network more dynamic range when predicting the frequency.
There is also a degeneracy between the ecliptic latitude and the frequency, therefore the network still needs access to the true frequency of the analysis band.
The parameter $f_{\mathrm min}$ is appended to the inputs the the network, exactly where this is appended is described in more detail in Sec.~\ref{networkdesign}.

Finally the four transformed parameters are normalised between 0 and 1 such that we predict the four parameters,
\begin{equation}
\label{parameterisation:transform}
	\begin{split}
		p_1 &= \gamma/2\pi,  \\
		p_2 &= 2|\beta|/\pi, \\
		p_3 &=(f_0 - f_{\mathrm{min}})/(f_{\mathrm{max}} - f_{\mathrm{min}}),\\
		p_4 & = (\log{\dot{f}_0} - \log{\dot{f}_{\mathrm{min}}})/(\log{\dot{f}_{\mathrm{max}}} - \log{\dot{f}_{\mathrm{min}}}),
	\end{split}
\end{equation}
 where $\gamma$ and $\beta$ are the ecliptic longitude and latitude, $f_0$ is the initial frequency, $\dot{f}_0$ is the frequency derivative, $f_{\mathrm{min}, \mathrm{max}}$ are the upper and lower edges of the analysis band and $\dot{f}_{\mathrm{min}, \mathrm{max}}$ are the prior ranges for the first frequency derivative. 
 Once samples are genrenated in the $p_1,p_2,p_3,p_4$ space, they are converted back into the four Doppler parameters using the inverse of the transformations in Eq.~\ref{parameterisation:transform}.

\section{Network design \label{networkdesign}}

There are two distinct \ac{CVAE} structures shown in Tab.~\ref{networkdesign:table} which correspond to the two different data-sets described in Sec.~\ref{data}. 
This was required due to the vastly different noise distribution in the Viterbi tracks compared the the additive Gaussian noise case.
The latter case was used for development of the algorithm design and is not representative of the highly correlated noise that we observe in practice in Viterbi tracks output from the SOAP algorithm. 
The models used for the analysis contain the two encoders which approximate the distributions $r_{\theta_1}(z| y)$ and $q_{\phi}(z| x, y)$ and the decoder which approximates $r_{\theta_2}(x | z, y)$.
Each of which are composed of convolutional and fully connected layers and output some parameters which describe a probability distribution.
They each share a set of convolutional layers which aims to extract information that the three networks will share, this is then fed into separate fully connected layers. 
The weights and bias of the convolutional layers are shared between the networks as shown in Tab.~\ref{networkdesign:table}. 
The idea being that the representation of the $y$ data output from the convolutional layers should be common to all networks. 

The first of the \acp{CVAE} was designed for the Gaussian noise data-set, this the simpler of the two where $x$ is the re-parameterised Doppler parameters ($p_1,p_2,p_3$ and $p_4$) described in Sec.~\ref{data:parameterisation} and $y$ is the re-scaled frequency evolution described in Sec.~\ref{data:gaussnoise}.
Each of the encoders of this model output the means and standard deviations of $n_z$ independent Gaussian's, where $n_z$ is the size of the latent space $z$.
The size of the latent space $n_z$ was chosen to be 6 for this network, this was so that the latent space can encode information on at least the four Doppler parameters.
There was no improvement in increasing this value above 6 in tests of the networks structure.
The decoder network also outputs the means and standard deviations four independent Gaussian distributions corresponding to the four re-parameterised Doppler parameters. 

The parameters $x$ for the Viterbi \ac{CVAE} are the four re-parameterised Doppler parameters ($p_1,p_2,p_3$ and $p_4$) as well as the list of boolean values $b$ corresponding to the conditions that the track is associated with a signal, described in Sec.~\ref{data:viterbinoise}.
The inputs $y$ for the Viterbi network are the re-scaled Viterbi tracks described in Sec.~\ref{data:viterbinoise}.
As with the Gaussian noise \ac{CVAE}, the outputs of the two encoders are also the means and standard deviations of $n_z$ independent Gaussian distributions. 
In this case the latent space has a size $n_z = 128$, this was increased compared to the previous example as the number of inferred parameters $x$ has increased to from 4 to 366.
The goal was for the latent space to then learn information about each of the track conditions as well as the Doppler parameters, increasing the size of the latent space beyond $128$ did not improve the performance of the network, however this was not exhaustively tested. 
The outputs of the decoder are then the four Doppler parameters, which remain as the means and standard deviations of independent Gaussian distributions and the track conditions which are described by a probability of drawing a value of 1 from a Bernoulli distribution. 
The track conditions are included in this \ac{CVAE} to aid it in learning the Doppler posteriors more effectively, the posteriors we investigate in Sec.~\ref{results} are then marginalised over the track condition posteriors. 

\begin{table*}
	\caption{\label{networkdesign:table} This table show the network design for the two main networks used in this analysis. The covolutional layers span multiple columns as the weights are shared between the three networks. The outputs from the $r_{\theta_1}(z| y)$  and $q_{\phi}(z| x, y)$ networks are twice the number of latent space dimensions $z$ as they represent the means and log variances of a Gaussian distribution. Similarly the output of the $r_{\theta_2}(x | z, y)$ contains the means and log-variances of the four Doppler parameters. Quantities in square brackets are the output sizes from each of the layers, the quantities inside brackets for the Conv1D layers refer to the number of filters and the filter size respectively. The values inside the brackets for the Linear layers refers to the number of neurons used within that layers, the layer has this output size.}
	\bgroup
	\def\arraystretch{1.5}
	\centering
	\begin{tabular}{|c|| c| c| c|| c| c| c||}
		\hline
		Network & \multicolumn{3}{c||}{Gaussian noise network } &  \multicolumn{3}{c||}{Viterbi track network }\\
		\hline
		distribution & $r_{\theta_1}(z| y)$& $q_{\phi}(z| x, y)$ &$r_{\theta_2}(x | z, y)$ & $r_{\theta_1}(z| y)$&  $q_{\phi}(z| x, y)$ &$r_{\theta_2}(x | z, y)$ \\
		\hline
		Input sizes& \multicolumn{3}{c||}{$x=[4], \; y=[362,1], \; z=[6]$} & \multicolumn{3}{c||}{$x=[366], \; y=[362,1], \; z=[128]$}  \\
		\hline
		convolutional network & \multicolumn{3}{c||}{Conv1D(4, 4) [362,4]} & \multicolumn{3}{c||}{Conv1D(4, 4) [362,4]}  \\
		\hline
		 & \multicolumn{3}{c||}{ MaxPool(4) [90,4]} & \multicolumn{3}{c||}{ MaxPool(4) [90,4]}  \\
		\hline
		& \multicolumn{3}{c||}{Conv1D(4, 4) [90,4]} & \multicolumn{3}{c||}{Conv1D(4, 3) [90,4]}  \\
		\hline
		& \multicolumn{3}{c||}{ MaxPool(4) [90,4]} & \multicolumn{3}{c||}{ MaxPool(4) [90,4]}  \\
		\hline
		Flatten & \multicolumn{3}{c||}{ Flatten [360]}  & \multicolumn{3}{c||}{ Flatten  [360])} \\
		\hline
		Concatenate & Flatten [360]& Flatten + x [364]& Flatten+z [372] & Flatten [360] & Flatten + x [726]& Flatten+z [372] \\
		\hline
		Fully connected & Linear(64) & Linear(64) & Linear(64)  & Linear(64) & Linear(64) & Linear(64)   \\
		\hline
		 & Linear(64) & Linear(64) & Linear(64) & Linear(64) & Linear(64) & Linear(64)    \\
		\hline
		 Output & Linear(12) & Linear(12) & Linear(8)   & Linear(12) & Linear(12) & Linear(362 + 8)    \\
		\hline
	\end{tabular}
	\egroup
\end{table*}

\section{Training \label{training}}

\begin{figure}
	\includegraphics[width=\columnwidth]{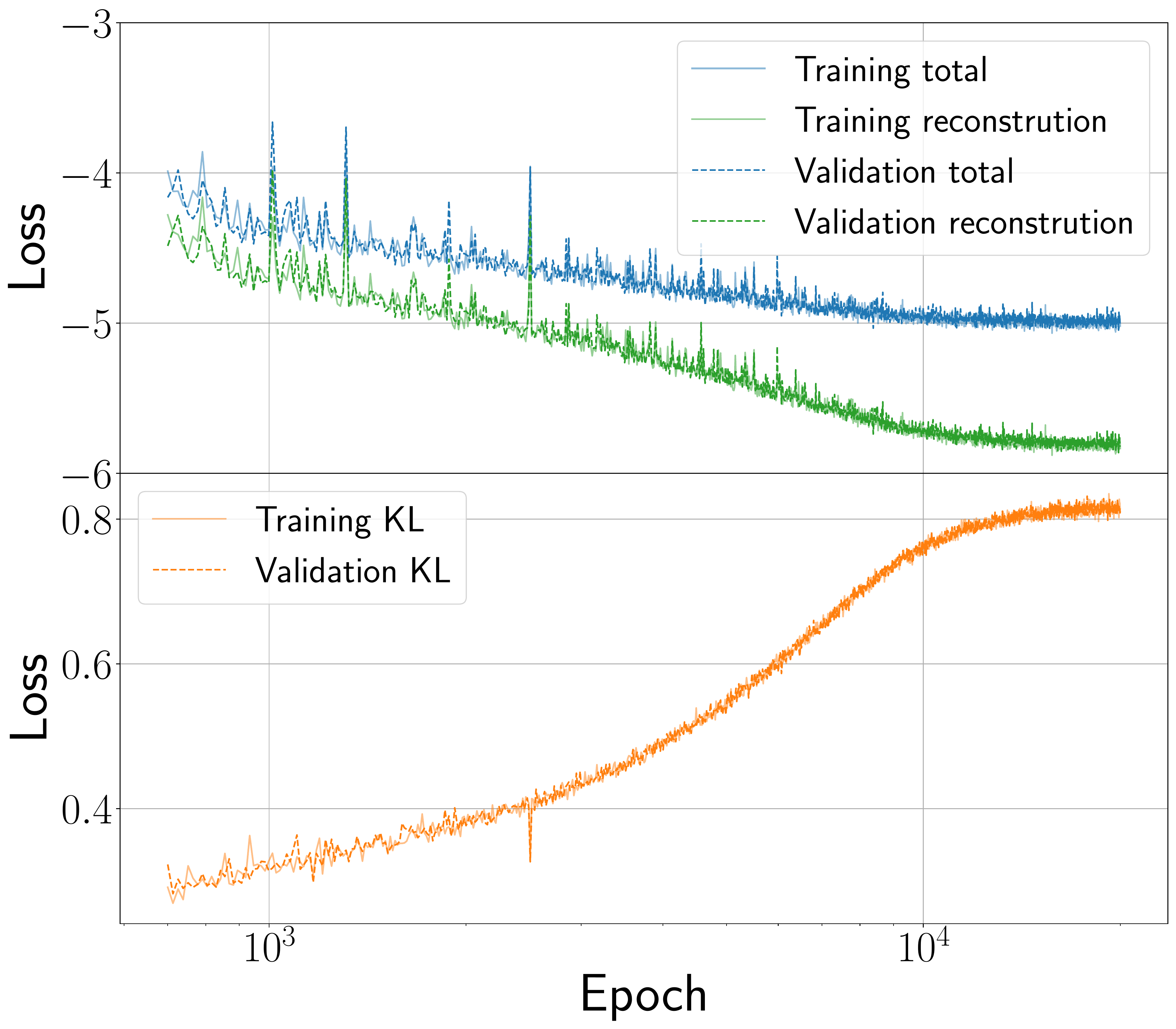}
	\caption{\label{cwgauss:losscurve} The total loss of the \ac{CVAE} (blue) is comprised of the reconstruction loss (green) and the KL divergence (orange). This is an example of a typical loss curve when trained on the Gaussian noise data-set described in Sec.~\ref{data:gaussnoise}. The plot begins at epoch 700 as during the ramping stage of training described in Sec.~\ref{ml:training} the loss reaches large values making the loss at later epochs difficult to read. }
\end{figure}

The training procedure involves splitting the training data into batches of 500, one batch of 500 signals and parameters are propagated through the \ac{CVAE}, where the average loss is calculated, i.e. the cost value in Eq.~\ref{cvae:cost}. This cost is then used to update the weights of the networks via back propagation. 	
This process is repeated for all of the batches in the training data where once the \ac{CVAE} has seen all of the training data, one epoch is complete. This process is repeated for 20000 epochs, where the weights are updated a small amount over each batch, therefore, the overall cost slowly moves towards a minimum. 
We slowly ramp up the influence of the KL divergence term in the loss, this is a linear ramp from 0 to 1 between the epochs 600 and 900.
We also apply a decay in the learning rate, where we multiply the learning rate by 0.993 every 5 epochs starting at epoch 4000. 
The values for the linear ramp and the decay rate were chosen such that the network performance improved, however they were not exhaustively optimised.
An example of the training and validation loss curves are shown in Fig.~\ref{cwgauss:losscurve}, this shows six curves corresponding to the total cost in Eq.~\ref{cvae:cost}, the reconstruction cost $L$ and KL-Divergence cost $KL$.
This show evidence that the \ac{CVAE} is not over-fitting to the training set as the validation and training loss curves overlap through the training. 
Also as the total loss curve (blue curve in upper panel of Fig.~\ref{cwgauss:losscurve}) appears to no longer be decreasing towards the end of training, this implies that the network has converged on a result.

\section{Timing \label{timing}}
One of the key focuses of the SOAP search is its ability to rapidly return results, therefore, this method should not drastically increase this time.
For $10^6$ training examples and 20000 epochs of training, it takes $\sim 5$ days to train the network using a Nvidia TITAN X GPU, this can now be sped up drastically by using more modern GPUs.
Whilst the training time is significant compared to the run time of SOAP, the training is completed only once before the search is completed. 
To generate 5000 samples from the 366 dimensional posteriors for the 400 test signals described in Sec.~\ref{data:viterbinoise} it takes a total 24s on the same Nvidia TITAN X GPU, leaving an average time to generate a posterior of 0.06s. 
Therefore, this does not add any significant time to the SOAP search.

\section{\label{results} Results}
%

To test our implementation of a \ac{CVAE} described in Sec.~\ref{machine} we use the two data-sets described in Sec.~\ref{data}.
The first data-set is used such that we can show a direct comparison between this method and traditional Bayesian sampling methods such as Dynesty \cite{speagle2019DynestyDynamic}.
The second data-set is a realistic example of the data which will be analysed in a real search using SOAP \cite{bayley2019SOAPGeneralised}.

\subsection{Gaussian noise}
\label{results:gaussnoise}

\begin{figure}
	\includegraphics[width=\linewidth]{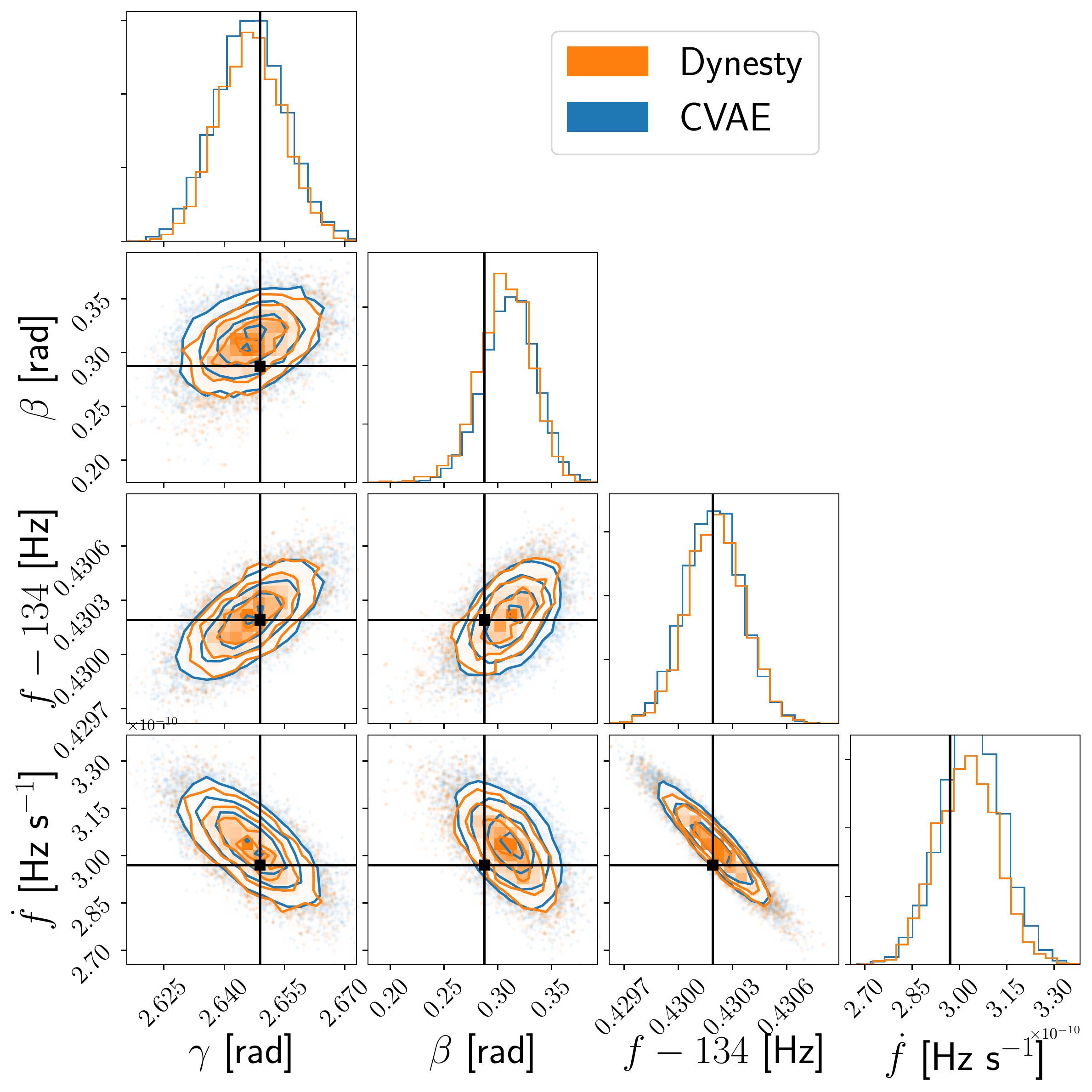}
	\caption{\label{cwgauss:posterior} An example posterior for a single frequency track with additive Gaussian noise. Blue is the \ac{CVAE} posterior and orange is the posterior from dynesty. The black markers who the true injection parameters. }
\end{figure}

To test the \ac{CVAE} 400 pieces of data are generated using the same methods as outlined in Sec.~\ref{data:gaussnoise}, this data-set is not used during the training procedure. 
Each of the pieces of test data are input to the \ac{CVAE} which then generates 10000 samples from the respective posterior distributions on the four Doppler parameters ($\gamma, \beta, f_0, \dot{f}_0$).
We then run the nested sampling algorithm Dynesty \cite{speagle2019DynestyDynamic} on the same pieces of test data, generating samples from the posterior on the same Doppler parameters.
Dynesty is run using a Gaussian likelihood function with a fixed noise variance of 0.01Hz and 1000 live points.
To demonstrate the accuracy of the \ac{CVAE} we show the comparison of the posterior samples from the \ac{CVAE} and dynesty for each of the test examples.
Figure \ref{cwgauss:posterior} shows this from one piece of test data, where we can see strong agreement between Dynesty (blue) and the \ac{CVAE} (orange).

\begin{figure}
		\includegraphics[width=\linewidth]{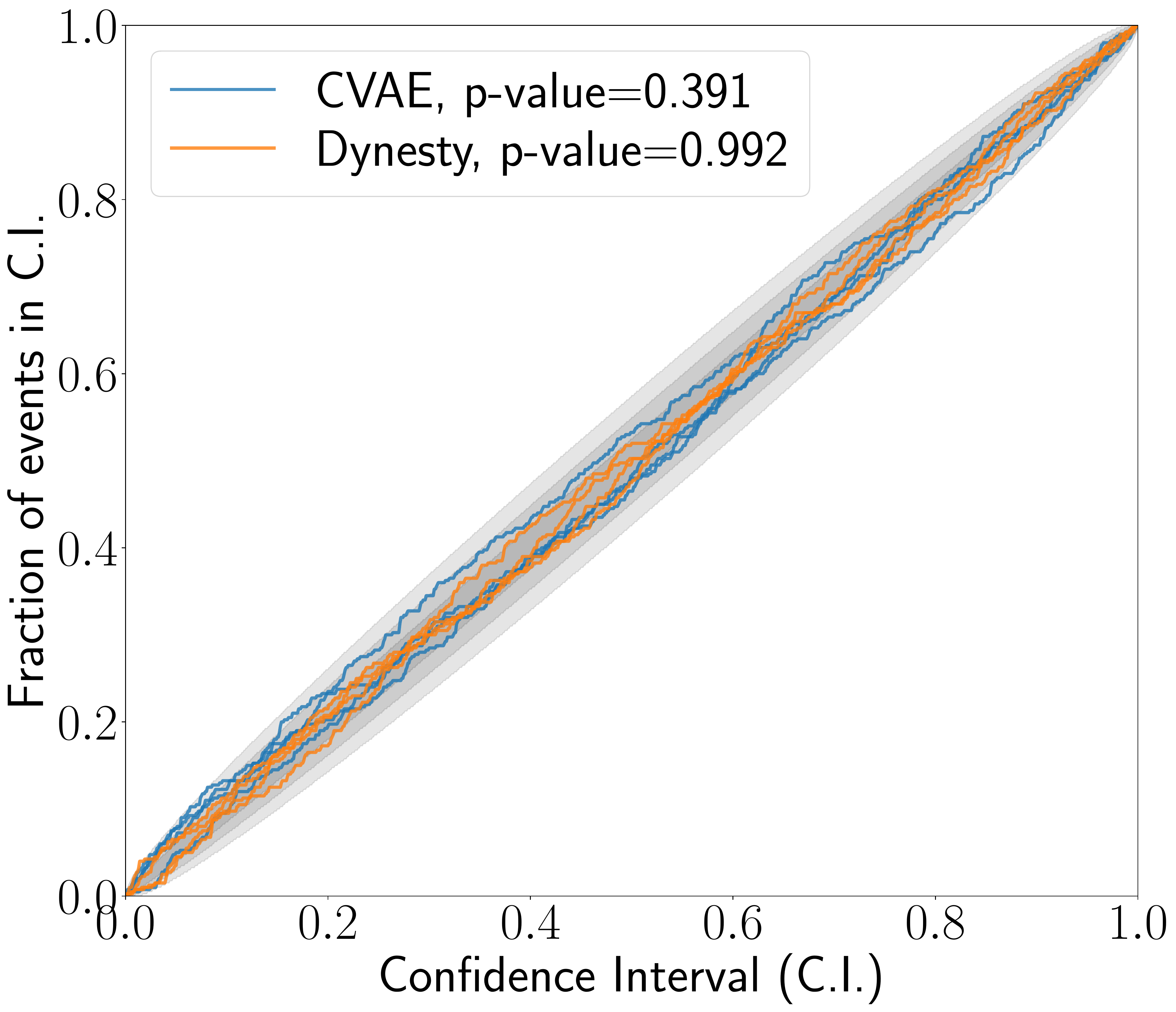}
		\caption{p-p plots are given for both samplers dynesty (orange) and our \ac{CVAE} implementation (blue). This is constructed from 400 test examples, where there are four curves for each of the two samplers corresponding to the four Doppler parameters. The grey regions refer to the one, two and three $\sigma$ confidence bounds expected from a uniform distribution for 400 test examples.}
		\label{cwgauss:ppplot}

\end{figure}

We can also run a statistical test over our entire test data-set by generating a probability-probability (p-p) plot.
A p-p plot is used to test that the posteriors are self-consistent and the true parameter values lie within the marginalised N\% confidence bounds for N\% of the simulations. 
If the methods are returning consistent posteriors then the p-p plot curve should be close to the diagonal. 
In Fig.~\ref{cwgauss:ppplot} we show the p-p plot of Dynesty compared with the p-p plot generated from the \ac{CVAE}. 
From these one can see that the \ac{CVAE} is consistent with the that from Dynesty.

\subsection{Viterbi noise}
\label{results:viterbi}

The main motivation for the work described in this paper was to estimate the posterior in the four Doppler parameters for a given Viterbi track.
To test how the \ac{CVAE} described in Sec.~\ref{networkdesign} performs on this task, a set of 500 Viterbi tracks are generated in the same way as Sec.~\ref{data:viterbinoise} and are not used in the training procedure.

\begin{figure*}
	\centering
    \includegraphics[width=\textwidth]{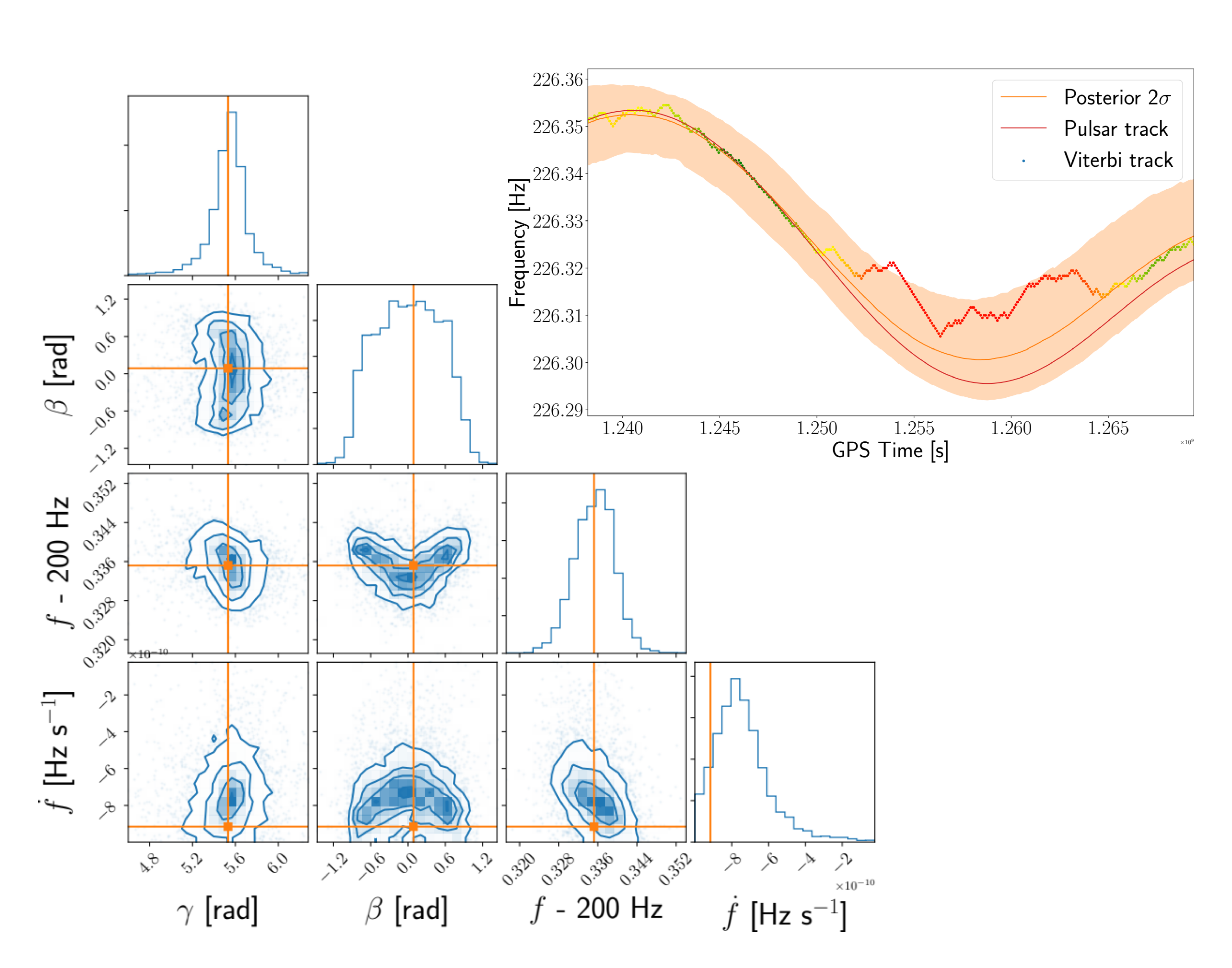}
	\caption{This shows the marginalised posterior distribution on the four Doppler parameters returned from a realistic Viterbi tracks, the true injected parameters are shown as the orange vertical and horizontal lines and the contours are at the 0.5,1,1.5 and 2 $\sigma$ level. In the top right a plot of the Viterbi tracks (green to red points) is also shown with the true pulsar frequency evolution (red) and a band containing tracks from the 2 sigma contours of the posterior (orange band). The red to green points of the Viterbi track correspond to the predicted probability that the track element is associated with a signal, green being more likely to be signal and red being more likely to be noise.}
	\label{cwsoap:all_example}
	
\end{figure*}

Each of the 500 Viterbi tracks are input to the \ac{CVAE} which then outputs 5000 samples from the posterior distribution on the four Doppler parameters and the 362 track conditions as described in Sec.~\ref{machine:testing}.
The marginalised posterior of the Doppler parameters are shown in Fig.~\ref{cwsoap:all_example} on the left hand side.
As we only generate samples from the posterior in the northern hemisphere of the sky, the $\beta$ posterior samples are reflected over the ecliptic equator ($\beta=0$) by randomly selecting half of the samples and inverting their sign.
The posterior for each of the track element probabilities are a set of binary samples drawn from different Bernoulli distributions.
For each time step the fraction of binary posterior samples that is equal to 1 is taken as a measure of the probability that the track is associated with the signal.
These fractions are represented in the upper right panel of Fig.~\ref{cwsoap:all_example}, where each sample of the Viterbi track is colored from red to green.
A track element colored green means that it is consistent with the signal and red means that it is consistent with noise. 

\begin{figure}
	\includegraphics[width = \linewidth]{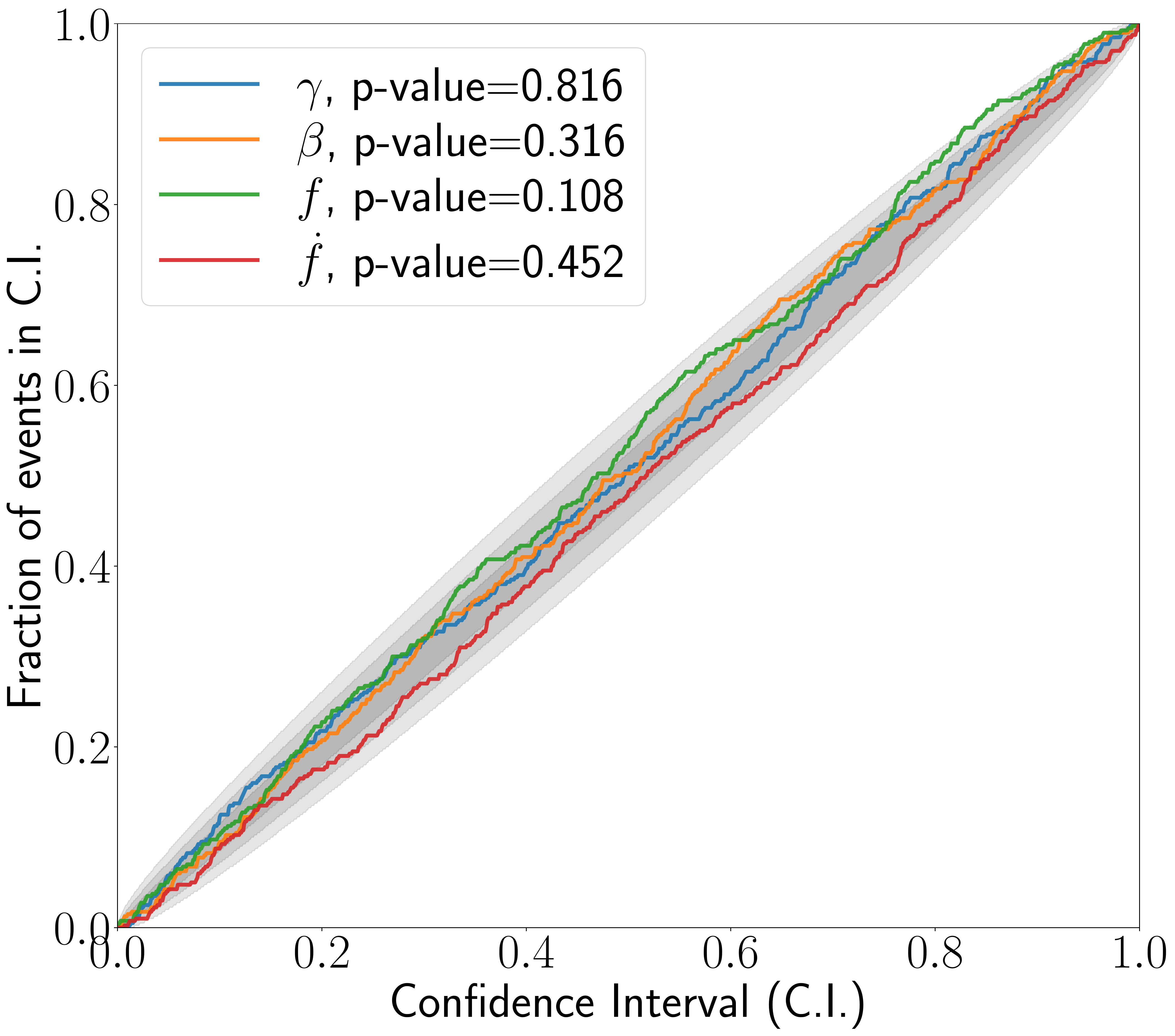}
	\caption{The p-p plot for the posteriors estimated by the \ac{CVAE} from a set of 500 Viterbi tracks. This is the fraction of events which lie in the given confidence interval of each of the 1D marginalised posterior distributions. The grey regions refer the the one, two and three $\sigma$ confidence bounds expected from a uniform distribution with 500 events. The combined p-value over all parameters is 0.36.}
	\label{cwsoap:ppplots}
\end{figure}

We are mainly interested in the Doppler parameters of the posterior and produce the posterior including the 365 track conditions mainly to assist the \ac{CVAE} in learning the distributions in the Doppler parameter space.
Therefore, for the majority of tests that follow we work only with the four dimensional marginal posteriors of the Doppler parameters. 
To test the consistency of the marginal posterior distributions on the Doppler parameters with the truths, we can generate a p-p plot for the four Doppler parameters as shown in Fig.~\ref{cwsoap:ppplots}. 
As described in Sec.~\ref{results:gaussnoise} the p-p plot shows that N\% of simulations lie within the N\% confidence reigon of the 1d marginalised posteriors.
Figure \ref{cwsoap:ppplots} shows that the p-p plot passes this test as the four curves remain within the 3$\sigma$ confidence bounds and a combined p-value of 0.36 is returned.
Whilst a p-p plot presents the effectiveness of the network on an ensemble of signals, it is also informative to see the performance on individual examples. 
Figure~\ref{cwsoap:all_example} shows an example output from generating a posterior on the Doppler parameters and track conditions using a \ac{CVAE}. 
This figure shows the marginalised posterior on the Doppler parameters on the left, demonstrating both that the Doppler parameters posterior is consistent with the injected parameter and that the \ac{CVAE} can reproduce more complex posteriors that in the previous test in Sec.~\ref{results:gaussnoise}.
In the upper right panel of Fig.~\ref{cwsoap:all_example}, one can see that the Viterbi track does not identify the entire signal, but around half way through the observation identifies noise instead.
The posterior conditions on the track elements effectively identify this region as originating from noise (colored red) and aids the \ac{CVAE} in generating Doppler posteriors more consistent with the truth.
Figure \ref{cwsoap:all_example} also shows a predicted frequency evolution over the Viterbi track using only the samples from the Doppler parameters.
The error bounds are are generated by taking the median and 90\% confidence interval of the track frequencies at each time step.

\subsubsection{Parameter space reduction}

The \ac{CVAE} returns a posterior distribution on the \ac{CW} Doppler parameters given an input Viterbi track, which in itself provides information on the source.
However, the main goal was to use this posterior to inform a more sensitive search such as a templated matched-filter search \cite{JKS,pyfstat}.
This would allow for easier verification of the source and would return more information on the Doppler parameters as well as other parameters associated with a \ac{CW}.
The matched-filter searches however, cannot be run over the entire Doppler parameter space as the number of templates required for an entire observing run would make the search computationally impossible.
Therefore, the reduction of the size of the parameter space using SOAP and the followup \ac{CVAE} is key for any follow-up search.
We can investigate what reduction in parameter space can be expected by applying this method compared to using just the SOAP search alone.

\begin{figure}
	\begin{subfigure}[b]{0.5\textwidth}
		\includegraphics[width = \textwidth]{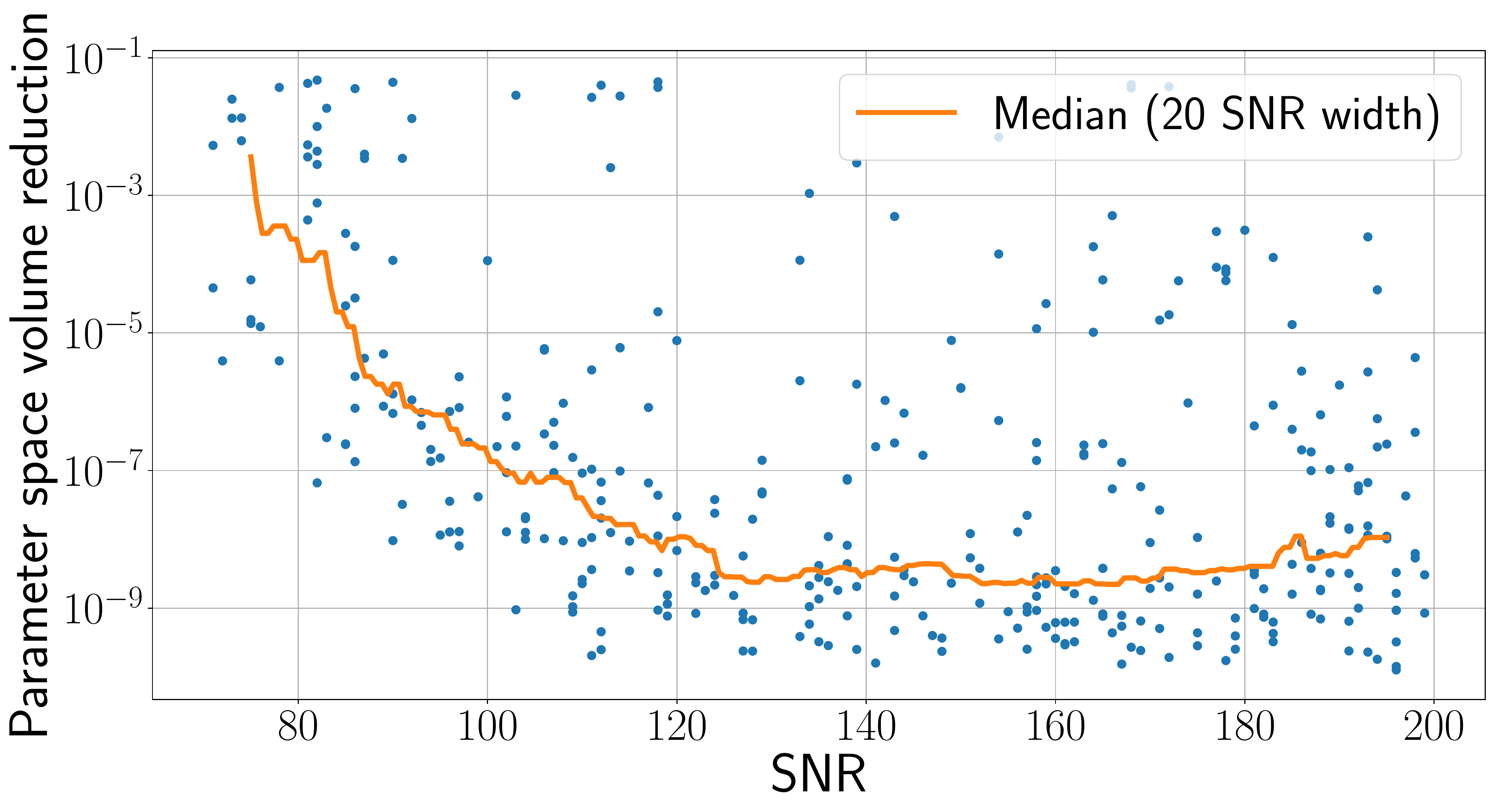}
		\caption{}
		\label{volume:reduction}
	\end{subfigure}

	\caption{The first panel shows the ratio of the volume contained within the 95\% contour of the Doppler parameters posterior and the full parameter space volume as a function of \ac{SNR}, where the full parameter space volume ranges are shown in Eq.~\ref{results:parspace:volume}. The orange curve shows the running median with a width of 8 bins.}
	\label{cwsoap:parametervolume}
\end{figure}

For an all-sky search the entire parameter space volume of the Doppler parameters can be found by looking at the ranges in which we search.
After the SOAP search has run these parameters are limited to 
\begin{equation}
    \label{results:parspace:volume}
	\begin{split}
		&\gamma \in [0, 2\pi] \; \mathrm{rad}\\
		 &\beta \in [-\pi/2, \pi/2] \; \mathrm{rad}\\
		  &f_0 \in 0.1 \; \mathrm{Hz}\\
		  &\dot{f}_0 \in [-1 \times 10^{-11}, 0] \; \mathrm{Hz \; s}^{-1} ,
	\end{split}
\end{equation}
where $f_0$ is limited to the 0.1 Hz wide sub-band width searched over by SOAP.
For each of the test examples which cross the SOAP detection threshold, we can make an estimate of the parameter space volume which is contained within a 95\% confidence region, then compare this to the total search volume. 
Figure~\ref{cwsoap:parametervolume} shows the reduction in parameter space as a function of the \ac{SNR} of a signal, this is the ratio of the volume contained in the 95\% region of the Doppler posterior compared to the total volume defined by the ranges in Eq.~\ref{results:parspace:volume}.
Figure~\ref{cwsoap:parametervolume} shows that at an \ac{SNR} of 100 the median that the parameter space is reduced is by a factor of $10^{-7}$.

\begin{figure}
	
	\begin{subfigure}[b]{0.5\textwidth}
		\includegraphics[width = \textwidth]{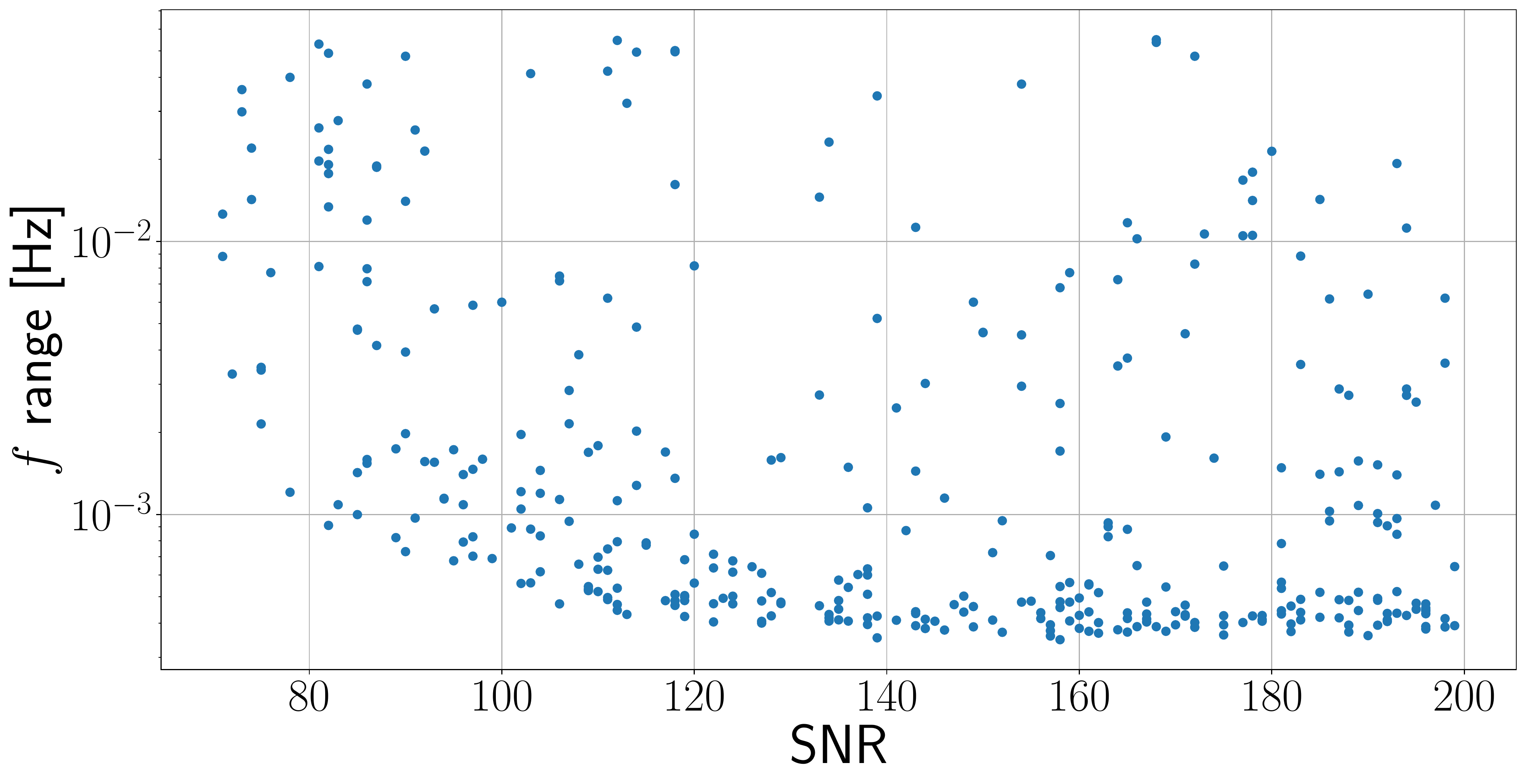}
		\caption{}
		\label{f:reduction}
	\end{subfigure}
	
	\begin{subfigure}[b]{0.5\textwidth}
		\includegraphics[width = \textwidth]{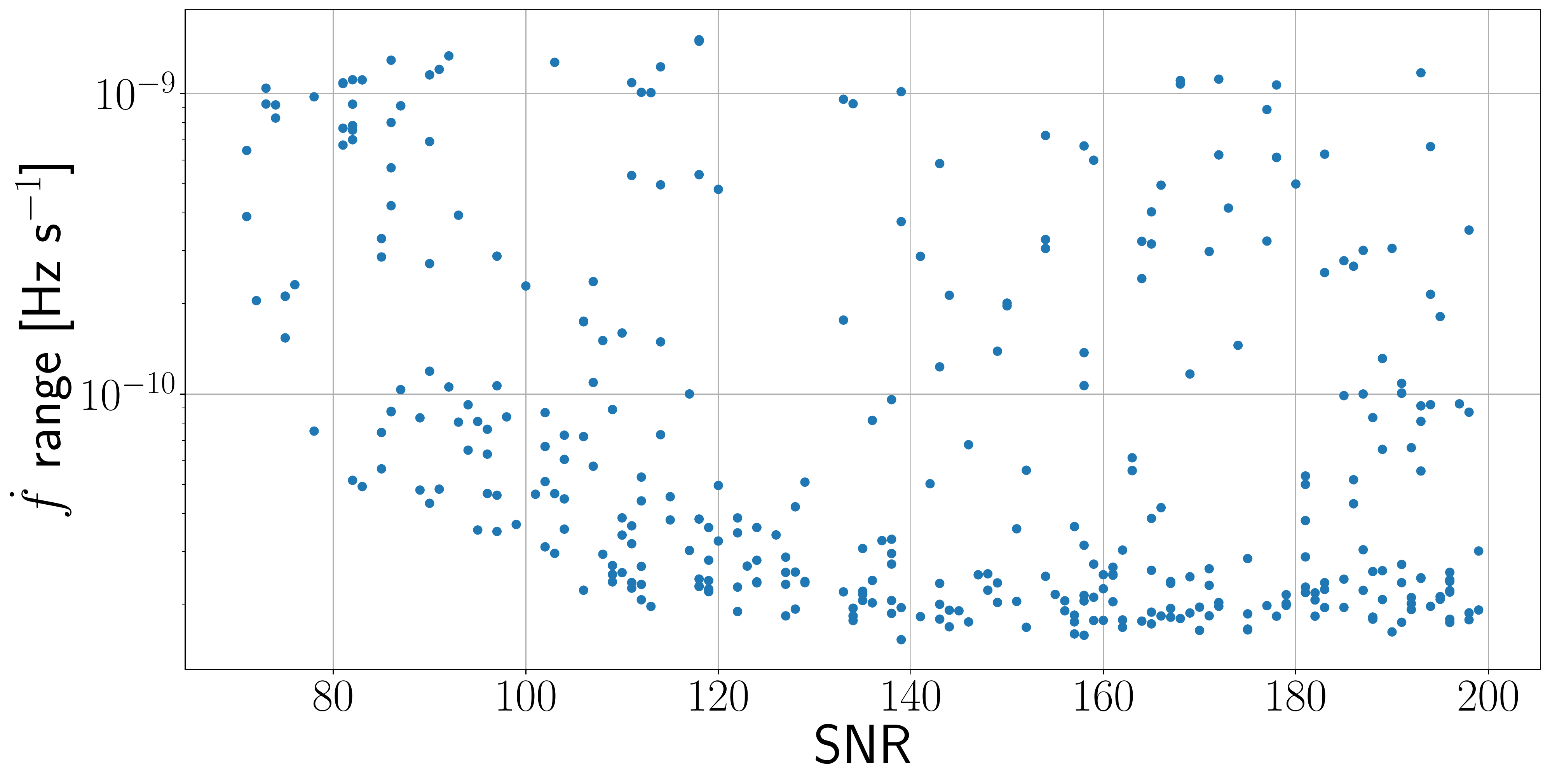}
		\caption{}
		\label{fdot:reduction}
	\end{subfigure}
	
	\begin{subfigure}[b]{0.5\textwidth}
		\includegraphics[width = \textwidth]{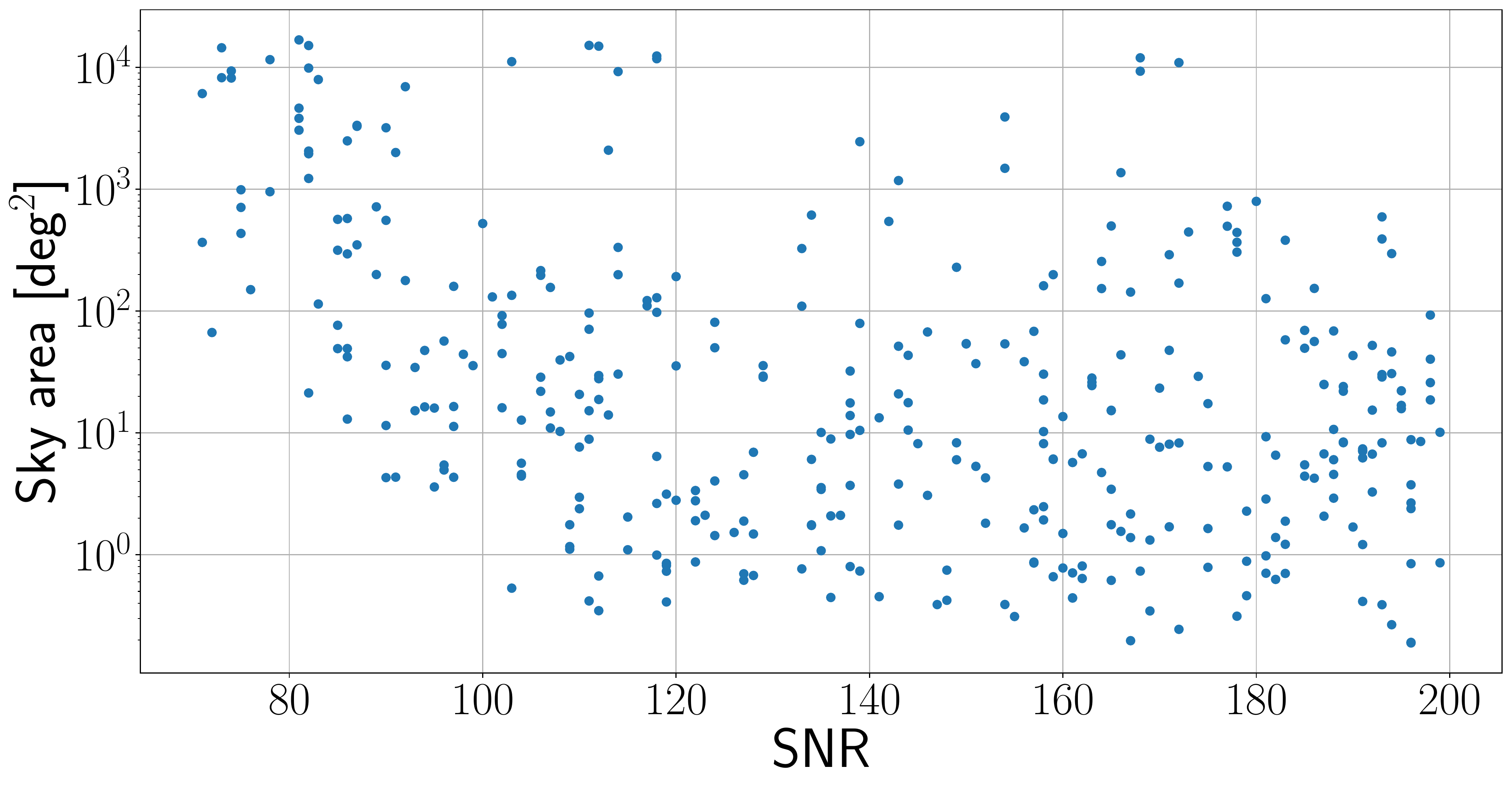}
		\caption{}
		\label{parvolume:reduction}
	\end{subfigure}

	\caption{ The first panel shows the frequency range contained within the 95\% intervals of the marginalised $f$ posterior. The second panel shows the frequency derivative range contained within the 95\% intervals of the marginalised $\dot{f}$ posterior. The final panel shows the sky area contained within the 95\% contour of the posterior on the sky parameters as a function of \ac{SNR}.}
	\label{cwsoap:parameterreds}
\end{figure}

In Fig.~\ref{cwsoap:parameterreds} the size of the region contained with 95\% of the marginal posteriors for the two frequency parameters and the sky position is shown.
When the signal has low \ac{SNR} SOAP identifies less of the Viterbi track and therefore there is not as much information in the track to help this follow-up to reduce the parameter space, leading to large parameters space regions at low \ac{SNR} in Fig.~\ref{cwsoap:parameterreds}.
There is also a large spread on the parameter regions for all of the parameters even for higher \acp{SNR}.
This can also be associated with SOAP not identifying the entire track, which occur if the signal drifts outside of the 0.1 Hz wide search band.
This is the case for many of the high \ac{SNR} large parameter space region points in Fig.~\ref{cwsoap:parameterreds}.

Due to the small reduction in parameter space in some of these signals, not all follow up methods will be suitable for all the signals.
What is more likely is that for each of the signals either a hierarchical semi-coherent approach would be used which is more sensitive than SOAP. 
This would include searches based on matched filters \cite{JKS,pyfstat,Tenorio.followup} or other semi-coherent methods such as \cite{SkyHough,FreqHough}.
For a hierarchical search, this method could act as a rapid initial stage of the search,  where the choice of length of the coherent segment would depend on the size of the posterior, i.e. longer coherence times can be used with smaller posteriors.

\section{Summary}

In this paper we describe a method to extract the source parameters of a \ac{CW} signal from the outputs of SOAP \cite{bayley2019SOAPGeneralised, bayley2020Soapcw}, which is an all-sky search for weakly modelled \acp{CW}.
This would allow for a more sensitive but more computationally expensive follow up search to use this narrower parameter space.
The paper outlines the machine learning methods which were used to extract these parameters and presents results from a number of tests of the validity of the outputs.

The outputs of the SOAP search include the time-frequency evolution of a candidate signal, which can randomly wander through a frequency band producing tracks which are highly correlated and difficult to define a likelihood for.
Traditional sampling methods cannot be used for this particular problem as a clear way to calculate the likelihood is required.
We therefore used likelihood free methods to extract the Bayesian posteriors, in particular we used a form of \ac{CVAE}. 
This allows us to extract Bayesian posteriors without ever being trained on the true posteriors. 
We outline this method and describe adaptations which were required when testing on two different datasets.

We test the method in two different simulated data-sets, a \ac{CW} frequency evolution with Gaussian noise added to the frequency bin locations and Viterbi tracks generated from \ac{CW} signals injected into Gaussian noise time series.
This allows us to compare the \ac{CVAE} approach to traditional sampling methods as well as demonstrate its performance in a realistic simulation.
When tested in the unrealistic data with Gaussian noise added to the frequency locations, the structure of the \ac{CVAE} is the simpler of the two models with its output being samples from the posterior of the 4 Doppler parameters.
In this simplified case, we show that the \ac{CVAE} can return a posterior which is consistent with one returned from a nested sampling method (dynesty).
As well as this we show a p-p plot which shows how the posteriors are statistically self consistent and are consistent with the simulated parameters.

The \ac{CVAE} was also tested with Viterbi tracks output from the SOAP search.
When testing on Viterbi tracks the \ac{CVAE} was modified such that it output not only posterior samples of the four Doppler parameters but also binary posterior samples from the conditions that the track element is associated with the true astrophysical signal.
This also allows us to infer which areas of the Viterbi track are associated with the signal.
Traditional sampling methods cannot be used with the Viterbi tracks as we have no clear definition of the likelihood, therefore we do not have a direct comparison between posteriors as in the previous test.
To test the output we instead demonstrated that the posterior distributions are statistically self consistent and consistent with the true parameters using a p-p plot.

The main motivation for this method as an addition to SOAP was to reduce the parameter space for a follow-up search using a more sensitive algorithm.
To asses the ability of the entire method to reduce the parameter space, we show the size contained within 95\% of the posterior of each of the individual Doppler parameters and the total reduction in the parameters space as a function of signal \ac{SNR}. 
The median of the reduction of the parameters space at an \ac{SNR} of 100 is $\mathcal{O}(10^{-7})$, for higher \ac{SNR} signals this reduces closer to $\mathcal{O}(10^{-9})$.
For low \ac{SNR} signals near the detection threshold, the median reduction in the parameter space is closer to $\mathcal{O}(10^{-3})$ which is expected as SOAP identifies less of the true signal at lower \ac{SNR}.

This method then extends the ability of the SOAP search allowing it to provide useful outputs to follow-up searches.
Now SOAP does not only rapidly search through large quantities of data returning likely long duration \ac{GW} candidates within $\mathcal{O}$(hour), but also rapidly returns Bayesian posteriors on the Doppler parameters of the identified signal in less that $\mathcal{O}(10^{-1})$ seconds per candidate.

\section{Acknowledgements}
We would like to acknowledge the continuous wave working group of
\ac{LIGO}-Virgo-KAGRA Collaboration for their assistance during this project. This research 
is supported by the Science and Technology Facilities Council., J.B.\, G.W.\ and C.M.\ 
are supported by the Science
and Technology Research Council (grant No. ST/V005634/1).
C.M.\ is also supported by the European Cooperation in Science and Technology
(COST) action CA17137. The authors are grateful for computational resources
provided by the LIGO Laboratory supported by National Science Foundation
Grants PHY-0757058 and PHY-0823459.

\bibliography{soap_parameter}

\end{document}